\documentclass[a4paper,11pt]{article}
\usepackage[utf8]{inputenc}
\usepackage[left=2.5cm, right=2cm, top=2.5cm, bottom=2cm]{geometry}
\usepackage[T1]{fontenc}
\newcommand{\nought}[1]{n_\up(#1)}
\newcommand{\pos}[1]{P_\up(#1)}
\newcommand{\ePoly}[2]{e_{#1}(#2)}
\newcommand{\Ccoeff}[1]{C_{#1}(N)}
\usepackage{amsmath}
\usepackage{hyperref}
\usepackage{graphicx,bm,amsmath,color}
\usepackage{bbold}
\usepackage{wasysym}
\usepackage{verbatim}
\usepackage{amssymb}
\usepackage{epstopdf}
\usepackage{animate}
\usepackage{xmpmulti}
\usepackage{centernot}
\usepackage{multirow}
\usepackage{tikz-cd}
\usetikzlibrary{calc}
\usetikzlibrary{intersections}
\usetikzlibrary{shapes.geometric}
\usetikzlibrary{positioning}
\usepackage[graphicx]{realboxes}
\usepackage{float}
\usepackage{mathtools}
\usepackage{comment}
\usepackage{braket, tensor}
\usepackage{cite}
\usepackage{adjustbox}
\usepackage[normalem]{ulem}

\def\co{\Delta}  
\newcommand{\be}{\begin{equation}}

\newcommand{\ee}{\end{equation}}
\newcommand{\beq}{\begin{eqnarray}}
\newcommand{\eeq}{\end{eqnarray}}
\newcommand{\ba}{\begin{align}}
\newcommand{\ea}{\end{align}}

\newcommand{\up}{\uparrow}
\newcommand{\down}{\downarrow}

\begin{document}

\begin{center}
\baselineskip 24 pt 
{\LARGE \bf  
Non-standard quantum algebra  $\mathcal{U}_h (\mathfrak{sl}(2, \mathbb{R}))$ and \\
$h$-Dicke states}
\end{center}

\begin{center}

{\sc A. Ballesteros$^{1}$, J.J. Relancio$^{2,3}$ and L. Santamar\'ia-Sanz$^{2}$}

\medskip
{$^1$Departamento de Física, Universidad de Burgos, 
09001 Burgos, Spain}

{$^2$Departamento de Matem\'aticas y Computaci\'on, Universidad de Burgos, 
09001 Burgos, Spain}

{$^3$Centro de Astropart\'{\i}culas y F\'{\i}sica de Altas Energ\'{\i}as (CAPA),
Universidad de Zaragoza, Zaragoza 50009, Spain}
\medskip
 
e-mail: {\href{mailto:angelb@ubu.es}{angelb@ubu.es}, \href{mailto:jjrelancio@ubu.es}{jjrelancio@ubu.es},
\href{mailto:lssanz@ubu.es}{lssanz@ubu.es}}

\end{center}

\begin{abstract}
We discuss the application of the Jordanian quantum algebra $\mathcal{U}_h (\mathfrak{sl}(2, \mathbb{R}))$, a Hopf algebra deformation of the Lie algebra $\mathfrak{sl}(2, \mathbb{R})$, in order to generate sets of $N$ qubit quantum states. We construct the associated $h$-deformed Dicke states using the Clebsch-Gordan coefficients for $\mathcal{U}_h (\mathfrak{sl}(2, \mathbb{R}))$, showing that the former exhibit completely different features than the $q$-Dicke states obtained from the standard quantum deformation $\mathcal U_q (\mathfrak{sl}(2, \mathbb{R}))$. Moreover, the density matrices of these $h$-deformed Dicke states are compared to the experimental realizations of those of Dicke states, and several similarities are observed, indicating that the $h$-deformation could be used to describe noise and decoherence effects in experimental settings, as well as to control the degree of entanglement of the state in quantum computing protocols. In particular, $h$-Dicke states for $N=2,3,4$ are presented, a method to construct the $h$-deformed analogs of $W$-states for arbitrary $N$ is given, and some algebraic considerations for the explicit derivation of generic $h$-Dicke states are provided.

\end{abstract}\thispagestyle{empty}

\tableofcontents

\section{Introduction}

Quantum algebras are deformations of Hopf algebras~\cite{Abe} and can also be regarded as the Hopf algebra duals of quantum groups, which generalize classical Lie groups into the noncommutative domain~\cite{Chari,Majid}. These quantum algebras are constructed by deforming not only the commutation relations but also the coproduct (which governs how tensor product representations are formed) of the universal enveloping algebra of a given Lie algebra. This deformation is controlled by a quantum parameter~\cite{Drinfeld,Jimbo}. Originally introduced as new symmetry structures in (1+1)-dimensional quantum integrable field theories, these algebras have since found a wide range of applications in exactly solvable spin chains, nuclear physics, noncommutative geometry, and quantum gravity (see~\cite{Majid:1988we,GomezSierra,Chaichian,Bonatsos,Szabo:2001kg,QGP} and references therein). It is worth emphasizing that the interpretation of the deformation parameter varies depending on the specific physical context.

{We also recall that a precursor to quantum groups are $q$-deformed harmonic oscillators. The first one to be introduced was the Arik-Coon $q$-oscillator~\cite{Arik:1973vg}. Later on, the Biedenharn-Macfarlane $q$-oscillator~\cite{Biedenharn:1989jw,Macfarlane:1989dt} was found to provide the building block for the representation theory of quantum algebras by generalizing the usual boson mapping techniques to the $q$-deformed case, including many-body problems~\cite{Floratos:1990uf,Biedenharnbook,Bonatsos}.
Such irreducible representations of quantum algebras on Hilbert spaces provide new sets of quantum states that are smooth deformations of the known ones coming from Lie algebra theory, except in the case when $q$ is a root of unity. Applications of these $q$-deformed states include the description of anharmonic and nonclassical effects in quantum optics (see, for instance,~\cite{Artoni:1993zz,Nelson:1993yj,meng:2007}) and, in the context of quantum information theory, the study of bipartite entanglement for $q$-oscillator and $q$-spin systems~\cite{Berrada2011,Berrada2012}, the construction of quantum logic gates \cite{Altintas:2014}, the study of Schr\"odinger cat states with enhanced nonclassical properties \cite{Dey:2015hma}, the improvement of post-selected weak measurements together with stronger nonclassical effects  \cite{Mousavigharalari:2025}, and the construction and analysis of deformed Dicke states~\cite{li2015entanglement,Raveh_2024,Ballesteros:2025dbv}, which will be the main topic of this paper.} 



The cornerstone of the study of quantum algebras is the quantum deformation of the Lie algebra $\mathfrak{sl}(2, \mathbb{R})$, which serves as the prototype for this theory. There are two fundamentally different types of deformation. The first is the standard or quasitriangular deformation, denoted $\mathcal U_q (\mathfrak{sl}(2, \mathbb{R}))$, which was the earliest to be introduced~\cite{KulishReshetikhin} and has received the most attention, particularly in relation to its applications. {The second potential deformation of $\mathfrak{sl}(2, \mathbb{R})$ is the non-standard one, commonly known as the triangular or Jordanian deformation and}  represented as $\mathcal U_h (\mathfrak{sl}(2, \mathbb{R}))$, which will be the focus of the present work. Each of these deformations corresponds to a distinct constant solution of the Classical Yang-Baxter equation on $\mathfrak{sl}(2, \mathbb{R})$~\cite{Chari}. {We recall that both deformations can be superposed giving rise to the so-called `hybrid' $(q,h)$-deformation, which is nevertheless of the standard type since it can be obtained from the  standard $q$-deformation through a nonlinear change of basis either on the quantum group coordinates~\cite{Aghamohammadi:1994kk} or in terms of the quantum algebra generators~\cite{Ballesteros_1999}. In this context, the $h$-deformation can be understood as the $q\to 1$ limit of the hybrid $(q,h)$-deformation, but $q$ and $h$-deformations provide mathematically inequivalent Hopf algebras and representation theories, and will be considered separately.}

It is important to emphasize that the irreducible representations of the standard quantum algebra $\mathcal U_q (\mathfrak{sl}(2, \mathbb{R}))$ share a structural resemblance with those of the undeformed Lie algebra $\mathfrak{sl}(2, \mathbb{R})$, provided that the deformation parameter $q$ is not a root of unity~\cite{Pasquier}. However, this similarity does not extend to the non-standard quantum algebra $\mathcal U_h (\mathfrak{sl}(2, \mathbb{R}))$ or its associated quantum group~\cite{10.1143/PTPS.102.203, Zakrzewski,Ohn,AngelBallesteros_1996}. The finite-dimensional irreducible representations of $\mathcal U_h (\mathfrak{sl}(2, \mathbb{R}))$ have been investigated in~\cite{Dobrev, Abdesselam,Ballesteros1997,Aizawa1997}, and results concerning their tensor product decompositions and Clebsch–Gordan coefficients are available in~\cite{Aizawa1997,VanderJeugt}. In addition, infinite-dimensional representations have been analyzed in~\cite{Ballesteros1997,Aizawa1997}. One of the distinctive features of the $\mathcal U_h (\mathfrak{sl}(2, \mathbb{R}))$ deformation is its non-Hermitian character, which naturally raises questions about its relation to $\mathcal{PT}$-symmetry. For finite-dimensional cases, this issue was recently addressed in~\cite{Ballesteros_2024}, where $\mathcal{PT}$-symmetric Hamiltonians were constructed for general finite-dimensional irreducible representations of the non-standard $\mathcal U_h (\mathfrak{sl}(2, \mathbb{R}))$ quantum algebra, leading to a new class of exactly solvable models with real spectra. Its generalisation to infinite-dimensional representations was studied in \cite{ballesteros2025inf}. {Finally, we recall that the non-standard version of the $q$-oscillator was introduced in~\cite{BHosc_1996}, and its connection with finite range potentials were presented in~\cite{Ballesteros:2005px}.}


Among the most studied quantum states for an arbitrary number of qubits, a special interest has been put in  Dicke states~\cite{dicke1954coherence}, which are the fully symmetric eigenstates of the total spin operators in SU(2) theories. The Dicke states  are invariant under the permutation group. For a system of $N$ qubits with $k$ excitations, the corresponding Dicke state $\ket{D_N^k}$ in the undeformed theory reads
\begin{equation}
    \ket{D_N^k} := \sqrt{\frac{k! (N-k) !}{N !}}  \sum_l \mathcal{P}_l \left( \ket{\down}^{\otimes (N-k)} \otimes \ket{\up}^{\otimes k} \right),
    \label{dickes}
\end{equation}
where $\sum_l \mathcal{P}_l$ denotes the sum  over all possible permutations of the $N$ sites in the qubit chain, {and the notation 
\begin{equation}
  \left\{  \ket{\up} := 
\begin{pmatrix}
1 \\
0 \\
\end{pmatrix}, \qquad
    \ket{\down} :=
\begin{pmatrix}
0 \\
1 \\
\end{pmatrix}\right\}
\label{eq:basis}
\end{equation} is used\footnote{ {Throughout this article, we will use this choice of base for the finite dimensional Hilbert space $\mathbb C^{2}$.}}.} Owing to their symmetry, Dicke states are exceptionally valuable in quantum computing because they combine (i) strong multipartite entanglement with (ii) robustness against particle loss and (iii) metrological enhancement (see~\cite{Wieczorek2009,Prevedel2009,bengtsson2017geometry,walter2016multipartite,Marconi:2025ioa} and the references therein). In this work, we will focus on Dicke states and their analogs obtained through the representation theory of two different quantum deformations, comparing their properties and highlighting their possible applications.

This paper is organized as follows. Sec.~\ref{sec:revisiting} discusses the mathematical structure of the undeformed and deformed  $\mathcal U_q (\mathfrak{sl}(2, \mathbb{R}))$ and $\mathcal U_h (\mathfrak{sl}(2, \mathbb{R}))$ quantum algebras, including their commutation relations, coproduct maps, and representations. It also covers the construction of Dicke states in both undeformed and deformed settings, where Clebsch--Gordan coefficients will be essential in the deformed case. Sec.~\ref{sec:2q}~examines the explicit construction of two-qubit states under the $h$-deformation, analyze their properties (such as orthogonality and symmetry), and compares them with undeformed and $q$-deformed cases. This section also discusses Schmidt decomposition for these states. Sec.~\ref{sec:3q} extends the analysis to three-qubit systems, presenting the explicit forms of the $h$-deformed states, their symmetries, and classification according to the Ac\'{i}n scheme for multiparticle entanglement~\cite{acin2000generalized}. In Sec.~\ref{sec:gen} we provide an analytical expression for computing the $h$-Dicke states for up to 11 spins and for arbitrary spin systems. Finally, we end with some conclusions and further work in Sec.~\ref{sec:conclusions}.

\section{Revisiting  \texorpdfstring{$\mathcal U (\mathfrak{sl}(2, \mathbb{R}))$, $\mathcal U_q (\mathfrak{sl}(2, \mathbb{R}))$ and $\mathcal U_h (\mathfrak{sl}(2, \mathbb{R}))$}{k}  Hopf algebras}
\label{sec:revisiting}
The Lie algebra  $\mathfrak{sl}(2, \mathbb{R})$ is defined by the following commutation relations 
\begin{equation}
    [J_z,J_\pm]= \pm J_\pm,\qquad [J_+,J_-]= 2 J_z ,
    \label{eq:su2Lie}
\end{equation}
where 
\begin{equation}
    J_x = \frac{1}{2} \sigma_x, \qquad J_y = \frac{1}{2} \sigma_y, \qquad J_z = \frac{1}{2} \sigma_z, \qquad   J_\pm = J_x \pm i J_y ,
\end{equation}
gives the fundamental $2 \times 2$ finite dimensional irreducible representation in term of the Pauli matrices \cite{vanderWaerdenbook, Georgibook} 
\begin{equation}
\sigma_x = 
\begin{pmatrix}
0 & 1 \\
1 & 0 \\
\end{pmatrix}, \qquad
\sigma_y = 
\begin{pmatrix}
0 & -i \\
i & 0 \\
\end{pmatrix}, \qquad
\sigma_z = 
\begin{pmatrix}
1 & 0 \\
0 & -1 \\
\end{pmatrix}.
\end{equation}
The action of the spin operators $J$ on the basis vectors of the finite dimensional Hilbert space\footnote{A set of $N$ spins is quantum mechanically described by rays in the finite dimensional Hilbert space $(\mathbb C^{2})^{\otimes N}$.} $\mathbb C^{2}$ 
is given by 
\begin{equation}
 \begin{array}{ll}
  J_+ \ket{\down} = \ket{\up}, & \qquad J_+ \ket{\up} = 0,\\
  J_- \ket{\down} = 0, &\qquad J_- \ket{\up} = \ket{\down},\\
  J_z \ket{\down} = -\frac{1}{2} \ket{\down}, &\qquad J_z \ket{\up} = \frac{1}{2} \ket{\up}.
  \end{array}\label{genaction}
\end{equation}

The Hopf algebra structure of the universal enveloping algebra $\mathcal U (\mathfrak{sl}(2, \mathbb{R}))$ is given by the commutation relations \eqref{eq:su2Lie} together with the coproduct map $\Delta_0 : \mathcal U (\mathfrak{sl}(2, \mathbb{R})) \to \mathcal U (\mathfrak{sl}(2, \mathbb{R})) \otimes \mathcal U (\mathfrak{sl}(2, \mathbb{R}))$ defined as
\begin{equation}
   \Delta_0 (J_\pm)= 1 \otimes J_\pm+ J_\pm \otimes 1,\qquad \Delta_0 (J_z)= 1 \otimes J_z+ J_z \otimes 1 ,
\label{eq:cop0}
\end{equation}
a counit $\epsilon : \mathcal U (\mathfrak{sl}(2, \mathbb{R}))  \to \mathbb{C}$, and an antihomorphism called antipode $\gamma : \mathcal U (\mathfrak{sl}(2, \mathbb{R}))  \to  \mathcal U (\mathfrak{sl}(2, \mathbb{R})) $ such that:
\begin{eqnarray}
    \epsilon(J_z)= \epsilon(J_\pm)=0, \quad \epsilon(1)=1, \quad \gamma(J_z)=-J_z, \quad \gamma(J_\pm)=-J_\pm, \quad \gamma(1)=1.
\end{eqnarray}
Notice that the coproduct is a coassociative map, \emph{i.e.} $(\Delta \otimes \mathrm{id}) \Delta = (\mathrm{id} \otimes \Delta) \Delta$ (see more properties in \cite{Biedenharnbook}).

Two-particle states in $\mathfrak{sl}(2, \mathbb{R})$ can be obtained in the usual way \cite{Sakuraibook} through irreducible representations of $\mathfrak{sl}(2, \mathbb{R})$ given by
\begin{eqnarray}\label{boots1}
    \ket{j,m} = C^{j_1,j_2,j}_{m_1,m_2,m} \ket{j_1,m_1}\otimes \ket{j_2, m_2},
\end{eqnarray}
where $j=(j_1 + j_2)\,\dots,|j_1-j_2|$, $m_i=-j,...,j$, and $C^{j_1,j_2,j}_{m_1,m_2,m}$ are the associated Clebsch--Gordan coefficients
\footnotesize
\begin{eqnarray}\label{boots2}
    C^{j_1,j_2,j}_{m_1,m_2,m}&=&\delta_{m_1+m_2,m} \left(\frac{ (j_1+j_2-j)!
(j_1-j_2+j)! (-j_1+j_2+j)! }{(j_1+j_2+j+1)!} \right)^{1/2} \nonumber\\
&\times &\left((j_1+m_1)!(j_1-m_1)!(j_2+m_2)!(j_2-m_2)!(j+m)!(j-m)!(2j+1)
\right)^{1/2} \nonumber\\
&\times & \sum_k \frac{(-1)^k}{k!(j_1+j_2-j-k)!(j_1-m_1-k)!(j_2+m_2-k)!(j-j_2+m_1+k)! (j-j_1-m_2+k)!}.\nonumber\\
\end{eqnarray}
\normalsize
We will be interested in $N$-particle states arising from the composition of $N$ irreducible representations of spin-$1/2$ particles (qubits), which can be obtained by iterating the previous procedure. In particular, Dicke states $\ket{D_N^k}$~\eqref{dickes} are the basis vectors for the highest-dimensional $j=N/2$ irreducible representation arising from the decomposition into irreducible blocks. In Hopf-algebraic terms, Dicke states  can be constructed in the form~\cite{Ballesteros:2025dbv}
\begin{align}
\ket{D_N^k}:=\mathcal{N}^N_k\left(\Delta^{(N)}_{0}(J_+)\right)^k\ket{\down\down\dots\down},
\end{align}
where the normalization constants read
\begin{equation}
    \mathcal{N}^N_k =\prod_{i=1}^k \frac{1}{\sqrt{(N+1-i) i}}= \frac{1}{k!} \frac{1}{\sqrt{\binom{N}{k}}}.
    \label{norDicke}
\end{equation}
Recall that, in general, for any coproduct map, if we denote $\co^{(2)}\equiv\co$, then the homomorphism $\co^{(3)}:A\rightarrow
A\otimes A\otimes A$ can be defined using either of the following two equivalent expressions
\be \co^{(3)}:=(1\otimes\co^{(2)})\circ\co^{(2)}=(\co^{(2)}\otimes 1)\circ\co^{(2)}.
\label{fj}
\ee
This can be generalized to any number $N$ of tensor products of $A$ either by the next
recurrence relation
\be
\co^{(N)}:=(1^{\,\otimes (N-2)}\otimes
\co^{(2)})\circ\co^{(N-1)},
\label{fl}
\ee
or,  equivalently, by
\be
\co^{(N)}:=(\co^{(2)}\otimes 1^{\,\otimes (N-2)})\circ\co^{(N-1)}.
\label{fla}
\ee

\subsection{The standard deformation}

Concerning deformations, the Hopf algebra $\mathcal U_q (\mathfrak{sl}(2, \mathbb{R}))$ is generated by $\{L_z,L_+,L_-\}$ and can be defined in terms of the following relations~\cite{Biedenharnbook} where $q\in  {\rm I\!R}^+$:
\begin{equation}
[L_z,L_\pm]= \pm L_\pm,\qquad [L_+,L_-]= \left[ 2 L_z\right]_q = \frac{q^{L_z}-q^{-L_z}}{q^{1/2}-q^{-1/2}} .
\label{suq2}
\end{equation}
Above, we have made use of the symbol for $q$-numbers
\begin{equation}
[n]_q\coloneqq  \frac{q^{n/2}-q^{-n/2}}{q^{1/2}-q^{-1/2}},
\label{eq:qnumber}
\end{equation}
so
\be
 \left[ 2 L_z\right]_q = \frac{q^{L_z}-q^{-L_z}}{q^{1/2}-q^{-1/2}}=
  2 L_z +o[q^2],
\ee
and we recover the $\mathfrak{sl}(2, \mathbb{R})$ Lie algebra in the $q\rightarrow1$ limit. The co-algebra structure for the $\mathcal U_q (\mathfrak{sl}(2, \mathbb{R}))$ algebra is generated by the deformed coproduct map
\begin{equation}
\Delta_q (L_\pm)= q^{-L_z/2}\otimes L_\pm+ L_\pm \otimes q^{L_z/2},\qquad \Delta_q (L_z)= 1 \otimes L_z+ L_z \otimes 1.
\label{qcop}
\end{equation}
The counit is the same as that in the undeformed case. Regarding the antipode, we can verify that
\begin{eqnarray}
    \gamma(L_z)=-L_z, \quad \gamma(L_\pm)=-q^{\mp1/2}L_+, \quad \gamma(1)=1.
\end{eqnarray}
Irreducible representations for $\mathcal U_q (\mathfrak{sl}(2, \mathbb{R}))$ when $q$ is not a root of unity are defined through the action of the algebra generators onto an state $|j ,m\rangle$ in the form
\begin{equation}
L_\pm |j_i, m_i\rangle\, :=\, \sqrt{[j_i\mp m_i]_q [j_i\pm m_i+1]_q}\,\, |j_i ,m_i\pm 1\rangle,\qquad  L_z |j_i, m_i\rangle=m_i |j_i, m_i\rangle.
\label{repsuq2}
\end{equation}
Moreover, the deformed Casimir operator for the algebra~\eqref{suq2} reads
\begin{equation}
C_q= L_- L_+ + [L_z]_q[L_z+1]_q=L_+ L_- + [L_z]_q[L_z-1]_q \, ,
\label{qcas}
\end{equation}
and the eigenvalues of $C_q$ in the representation states $ |j_i, m_i\rangle$ are just $[j_i]_q[j_i+1]_q$. 
Therefore, if $q$ is not a root of unity, the representation theory of $\mathcal{U}_q(\mathfrak{sl}(2, \mathbb{R}))$ is structurally the same as that of $\mathfrak{sl}(2, \mathbb{R})$ (see~\cite{Biedenharnbook}). In fact, irreducible components in the tensor product (given by the deformed coproduct map~\eqref{qcop}) of two $\mathcal U_q (\mathfrak{sl}(2, \mathbb{R}))$ representations are again obtained in terms of \eqref{boots1}, where now the $q$-Clebsch--Gordan coefficients (see~\cite{alvarez2024russian} and references therein) are used:
\begin{equation}
\begin{split}
{\cal C}_{m_1, m_2, m}^{j_1, j_2, j} (q)\,=&\,\delta _{m,m_1+m_2} q^{\frac{1}{2} (j_1 m_2-j_2 m_1)-\frac{1}{4} (-j+j_1+j_2) (j+j_1+j_2+1)}
\\
 &\sqrt{\frac{[2 j+1]_q  [j+m]_q! [j_2-m_2]_q![j+j_1-j_2]_q! [-j+j_1+j_2]_q! [j+j_1+j_2+1]_q! }{[j-m]_q! [j_1-m_1]_q!  [j_1+m_1]_q! [j_2+m_2]_q! [j-j_1+j_2]_q!  }} 
\\
&\sum _{n=0}^{\min (-j+j_1+j_2,j_2-m_2)} \frac{ [2 j_2-n]_q!  (-1)^{-j+j_1+j_2+n} q^{\frac{1}{2} n (j_1+m_1)} [j_1+j_2-m-n]_q! }{ [n]_q!  [j_2-m_2-n]_q!   [-j+j_1+j_2-n]_q!  [j+j_1+j_2-n+1]_q! } \, .
\end{split}
\label{eq:qcg}
\end{equation}

Regarding the construction of Dicke states in this $q$-deformation, we will denote the tensor product state of $N$ qubits with $k$ excitations ($\up$) as $\ket{n_1,\dots,n_k}$ where $n_i$ indicates the positions of each of the $k$ excitations within the $N$-th tensor product. Thus $n_i\in\{1, \dots, N\}$ while $i\in \{1, \dots, k\}$. In this way, the $q$-Dicke states can be written as~\cite{li2015entanglement,Ballesteros:2025dbv} (see~\cite{Raveh_2024} for the generalization in the case of qudits):
\begin{equation}
      \ket{D_N^k}_q = \sqrt{\frac{[k]_q! [N-k]_q !}{[N]_q !}} \,\sum_{\mathbf n} q^{-k\left( \frac{N+1}{4} \right) + \frac{1}{2} \sum_{i=1}^k n_i}  \ket{\mathbf n},
      \label{qdickeexplicit}
\end{equation}
where the sum runs over all the possible sets $\mathbf n =(n_1,\dots,n_k)$ describing the location of the $k$ excitations. These states can be obtained by making use of the deformed coproduct~\eqref{qcop}, namely
\begin{align}
\ket{D_N^k}_q:=\mathcal{N}^N_k(q)\left(\Delta_q^{(N)}(L_+)\right)^k\ket{D_N^0}_q,
\label{qdickep}
\end{align}
where $\mathcal{N}^N_k(q)$ are the corresponding normalization constants. In an analogous fashion, the $q$-Dicke states can be obtained using 
\begin{align}
\ket{D_N^k}_q:=\tilde{\mathcal{N}}^N_k(q)\left(\Delta_q^{(N)}(L_-)\right)^{N-k}\ket{D_N^N}_q.
\label{qdickem}
\end{align}
For this deformation, one finds that $\ket{D_N^N}_q=\ket{\up\up\dots\up}$ and $\ket{D_N^0}_q=\ket{\down\down\dots\down}$. In contrast, in the non-standard deformation, $\ket{D_N^0}_h$ will no longer be equal to $\ket{\down\down\dots\down}$, as we will see later. However, it is still true that $\ket{D_N^N}_h=\ket{\up\up\dots\up}$, since ${\cal C}^{j_1,j_2,j}_{n_1,n_2,m}(h)={\cal C}^{j_1,j_2,j}_{n_1,n_2,m}$ if $m=n_1+n_2$~\cite{VanderJeugt}. 

{We note that, as shown in~\cite{Altintas:2014}, a single qubit state can be  obtained from the action of two $q$-oscillators onto their ground states. Therefore, the previous $q$-Dicke states~\eqref{qdickeexplicit} for $N$ qubits can be achieved by using the realization of the coproducts of the operators $L_\pm$ within~\eqref{qdickep}-\eqref{qdickem} in terms of $2N$ $q$-oscillators acting on their $2N$ ground states. In this approach, it is again the deformed coproducts $\Delta_q^{(N)}(L_\pm)$ that provide the $q$-deformed coupling between different qubits which is characteristic of $q$-Dicke states, since the fundamental $j=1/2$ representation of $\mathcal U_q (\mathfrak{sl}(2, \mathbb{R}))$ coincides with the one of $\mathcal U (\mathfrak{sl}(2, \mathbb{R}))$.}

\subsection{The non-standard deformation}

On the other hand, the Hopf algebra $\mathcal U_h (\mathfrak{sl}(2, \mathbb{R}))$ is generated by $\{X,Y,H\}$, and can be defined in terms of the following relations~\cite{Ohn} 
\begin{equation}
    [H,X]= \frac{2}{h} \sinh(hX),\qquad [H,Y]= -Y \cosh(hX)-\cosh(hX) Y, \qquad [X,Y]= H,
    \label{eq:su2hLieold}
\end{equation}
where $h\in \mathbb{R}$ is the deformation parameter. The coproducts are given as follows:
\begin{eqnarray}
 \Delta_h(X)&=& X \otimes 1+1 \otimes X,\nonumber\\
 \Delta_h(Y)&=& Y \otimes e^{hX}+ e^{-hX} \otimes Y,\nonumber\\
 \Delta_h(H)&=& H \otimes e^{hX}+ e^{-hX} \otimes H.
\end{eqnarray}
Under the change of basis
\begin{equation}
    Z_+ = \frac{2}{h} \tanh \left(\frac{hX}{2}\right), \qquad  Z_- = \cosh \left(\frac{hX}{2}\right) Y \cosh \left(\frac{hX}{2}\right),
\end{equation}
the above commutators transform into the usual ones for $\mathfrak{sl}(2, \mathbb{R})$, as displayed in Eq. \eqref{eq:su2Lie}, providing $H=2J_z, \, Z_\pm = J_\pm$. The generators $\{H, Z_\pm \}$ act on the basis \eqref{eq:basis} in the same manner as the generators of $\mathcal U (\mathfrak{sl}(2, \mathbb{R}))$, that is, by means of \eqref{genaction}. Therefore, the entire effect of the deformation is encoded by the action on multipartite states.

The $\mathcal U_h (\mathfrak{sl}(2, \mathbb{R}))$ co-algebra structure is given by the coproduct map, which reads \cite{Aizawa1997}
\begin{eqnarray}
\Delta_h(H) &= &  H\otimes 1 + 1 \otimes H + 2H \otimes \sum_{n=1}^\infty
\left(\frac{hZ_+}{ 2}  \right)^n +  \sum_{n=1}^\infty
\left(- \frac{hZ_+}{ 2} \right)^n \otimes 2H,\nonumber\\
\Delta_h(Z_+)  &= & (1\otimes Z_+ + Z_+ \otimes 1)\left(\sum_{n=0}^\infty \left(-\frac{h^2}{4}\right)^n
 \; Z_+^n \otimes Z_+^n \right),\nonumber\\
 \Delta_h(Z_-)  &= & Z_- \otimes \sum_{n=0}^{\infty} (n+1) \left(\frac{h Z_+}{2} \right)^n + 
     \sum_{n=0}^{\infty}\; (n+1) \left(-\frac{h Z_+}{2} \right)^n \otimes Z_- \nonumber \\
 & &+ h\left(C_h-{\frac{H^2}{ 4}} \right) \otimes \sum_{m=1}^{\infty} 
     m \left(\frac{h Z_+}{2} \right)^m - 
     \sum_{m=1}^{\infty} m \left( -\frac{h Z_+}{2} \right)^m \otimes 
     h \left(C_h-{\frac{H^2}{ 4}}\right)     \nonumber \\
 & &+ \left({\frac{h}{ 2}}\right)^2 Z_+Z_-Z_+ \otimes 
     \sum_{k=2}^{\infty}  (k-1) \left(\frac{h Z_+}{2} \right)^k + 
     \sum_{k=2}^{\infty}  (k-1) \left(-\frac{h Z_+}{2} \right)^k \otimes 
     \left({\frac{h}{ 2}
     }\right)^2 Z_+Z_-Z_+,\nonumber\\
\label{eq:coproduct_j}
\end{eqnarray}
where $C_h= Z_+ Z_-+ H^2/4 -H/2$ is the Casimir element. In the limit $h \to 0$ we recover the primitive coproduct \eqref{eq:cop0}.  

The Clebsch--Gordan coefficients for $\mathcal U_h (\mathfrak{sl}(2, \mathbb{R}))$ (denoted by ${\cal C}^{j_1,j_2,j}_{n_1,n_2,m}(h)$) can be derived from those of $\mathcal U (\mathfrak{sl}(2, \mathbb{R}))$  by means of \cite{VanderJeugt}
\begin{eqnarray}\label{program1}
&&{\cal C}^{j_1,j_2,j}_{n_1,n_2,m}(h) = \sum_{m_1+m_2=m}
C^{j_1,j_2,j}_{m_1,m_2,m} A^{m_1,m_2}_{n_1-m_1,n_2-m_2},
\end{eqnarray}
where
\begin{eqnarray}\label{program2}
&&  A^{m_1,m_2}_{k,l} = a^{m_1,m_2}_{k,l} \frac{\alpha_{j_1,m_1+k}\alpha_{j_2,m_2+l} }{
\alpha_{j_1,m_1}\alpha_{j_2,m_2}},\nonumber\\
 &&   \alpha_{j,m}= \sqrt{\frac{(j+m)!}{(j-m)!}},\nonumber \\
 &&   a_{k,l}^{m_1,m_2}= (-1)^k 2^{-k-l} h^{k+l} (b_{k,l}^{m_1,m_2}-b_{k-1,l-1}^{m_1,m_2}), \\
 &&b^{m_1,m_2}_{k,l} = 
\left\{ 
 \begin{array}{ll}
\frac{(-2m_1-k)_l (-2m_2-l)_k}{ k!l!} & \hbox{if }k\geq 0\hbox{ and }
l\geq 0 , \nonumber\\[2mm]
0 & \hbox{otherwise}.
 \end{array} \right.
\end{eqnarray}
Note that the above $h$-Clebsch--Gordan coefficients generally yield non-normalized states.

In the following sections, we will compute the Dicke states for the non-standard deformation $\mathcal U_h (\mathfrak{sl}(2, \mathbb{R}))$ in order to compare both deformations. For that aim, we will use the notation $\ket{D_N^{\tilde{m}}}_h$, where $\tilde{m}$ is the eigenvalue of the $H=2 J_z$ operator in the corresponding $j=N/2$ irreducible representation, so $\tilde{m}=2k-N$. Similarly, in $\ket{D_N^{m}}_q$, $m$ is the eigenvalue of the $J_z$ operator in the corresponding $j=N/2$ irreducible representation, so $m=k-N/2$.

\section{2-qubit \texorpdfstring{$h$}{h}-states}
\label{sec:2q}
Consider the space resulting from the tensor product of two ($N=2$) particles with spin $1/2$. By recursively applying the operator $\Delta_h (Z_-)$ to the highest weight state $\ket{\up\up}$ one can systematically obtain all the new states $\ket{D_{N=2}^H}$, where the superindex refers to the eigenvalue of the $H$ operator associated with this state. Another way to compute these states is to use the Clebsch--Gordan coefficients for $\mathcal U_h (\mathfrak{sl}(2, \mathbb{R}))$ given by \eqref{program1} and \eqref{program2}. Through either of these two routes, and after normalization, we find the triplet of $h$-Dicke states:
\begin{eqnarray}
&&{\ket{D^{-2}_2}_h =\frac{1}{h^2+4} \left(4\ket{\down \down} -2h\ket{\up \down}+2h\ket{ \down\up} + h^2\ket{\up\up}\right),}	\nonumber\\
&&\ket{D^{0}_2}_h =	\frac{1}{\sqrt{2}} (\ket{\up \down}+\ket{ \down\up}),	\label{eq:2qubits}\\
&&\ket{D^2_2}_h =\ket{\up\up},\nonumber
\label{2h_dicke}
\end{eqnarray}
together with the singlet state
\begin{equation}
\ket{M^0_2}_h =\frac{1}{\sqrt{h^2+2}}\left( \ket{\up \down}-\ket{ \down\up}-h \ket{\up\up} \right),
\label{Mh}
\end{equation}
where we have used the notation shown in Fig. \ref{fig:two_spin_tree_DU} for labeling the states.
\begin{figure}[h!]
    \centering
    \begin{tikzpicture}[
        level distance=1.5cm,
        level 1/.style={sibling distance=4cm},
        every node/.style={font=\small, align=center},
        vertical child/.style={
            edge from parent path={(\tikzparentnode.south) -- (\tikzchildnode.north)}
        }
    ]
        \node {$\tfrac{1}{2}\;\oplus\;\tfrac{1}{2}$}
            child {
                node {$1$}
                child[vertical child] { node {$D$} }
            }
            child {
                node {$0$}
                child[vertical child] { node {$M$} }
            }
        ;
    \end{tikzpicture}
    \caption{Coupling tree for two spin-\(\tfrac12\) particles, showing \(J=1,0\) states, labeled as \(D\) and  \(M\), respectively.}
    \label{fig:two_spin_tree_DU}
\end{figure}

Note that the singlet ($M$) and triplet ($D$) subspaces are not orthogonal to each other ({\em i.e., } $\braket{M_2^0|D_2^0}_h\neq 0$), contrary to what occurs in the undeformed theory. These deformed states exhibit a structural pattern consistent with that presented in \cite{Ballesteros1997}. It basically consists on certain states that are undeformed $\{\ket{D^{0}_2}_h, \ket{D^2_2}_h\}$ 
and the remaining states are a superposition of the undeformed state with other basis states multiplied by factors of the deformation parameter. In general, we will call $h$-Dicke states the ones of the basis for the irreducible representation with the highest dimension, {\em i.e.} $(N+1)$. Notice that $\ket{D^{-2}_2}_h$ is no longer a fully symmetric state due to the minus sign in the term $\ket{\up\down}$.

There is a crucial difference between $h$-states and the $q$-Dicke states obtained from the standard deformation, in which the normalized singlet and triplet states read~\cite{li2015entanglement,Ballesteros:2025cia}
\begin{eqnarray}
 &&   \ket{D^{-1}_2}_q = \ket{\down \down},	\nonumber\\
&&\ket{D^{0}_2}_q =	\frac{1}{\sqrt{[2]_q}} (q^{1/4}\ket{\down\up} + q^{-1/4} \ket{\up \down}),	\\
&&\ket{D^1_2}_q =\ket{\up\up},\nonumber\\[2.5ex]
&&\ket{M^0_2}_q =\frac{1}{\sqrt{[2]_q}} (-q^{-1/4}|\down\up\rangle + q^{1/4} \ket{\up \down}).
\label{eq:2qqubits}
\end{eqnarray}
Here, the states $\ket{D^{0}_2}_q$ and $\ket{M^0_2}_q$ are deformed by introducing $q$-coefficients that monotonically change the weights of each component. 
This is no longer the case in the non-standard deformation, where the coefficients of the linear combination of basic vectors in $\ket{D_N^{\tilde{m}}}$ in \eqref{eq:2qubits} and \eqref{Mh} have completely different functional forms. Notice that now, all the states in \eqref{eq:2qubits} and \eqref{Mh} are eigenvectors of  $\Delta(H)\ket{D_N^{\tilde{m}}} = \tilde{m}\ket{D_N^{\tilde{m}}}$. We stress  that the two triplet states
$\{\ket{D^{0}_2}_h, \ket{D^{2}_2}_h\}$  are  undeformed Dicke states, while  $\ket{D^{-2}_2}_h$ is $h$-deformed and invariant under the exchange $h \to - h$ together with the permutation $\ket{ ik}\to \ket{ ki}$.  In contrast, the singlet state $\ket{M^0_2}_h$ is also invariant (up to a global sign) under the same transformation. We recall that  symmetry  properties are essential to characterize the space of states, and in the case of $q$-Dicke states, the corresponding $q$-permutational symmetries have been studied  in~\cite{Ballesteros:2025cia}.

\begin{figure}[ht]
    \centering
    \includegraphics[scale=0.75]{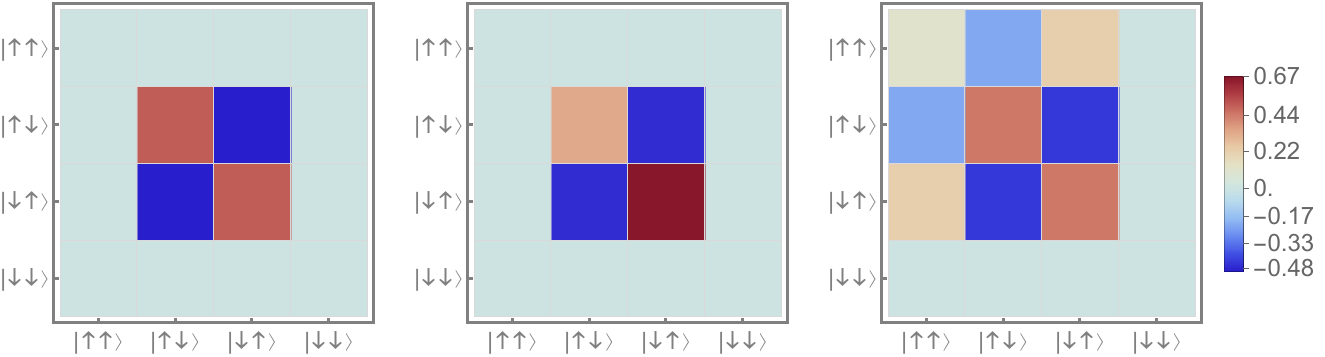}\\[2ex] \includegraphics[scale=0.65]{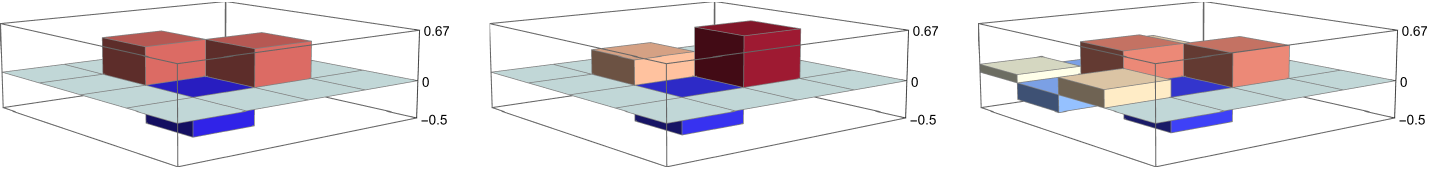}
    \caption{Left: Density matrix for the undeformed Bell state $\ket{\Psi^-}= (\ket{\up\down}-\ket{\down\up })/\sqrt{2}$ in the computational basis. Middle: Corresponding density matrix of the $q$-deformed state $\ket{M^0_2}_q$ for $q=0.5$. Right: Corresponding density matrix of the $h$-deformed  state $\ket{M^0_2}_h$ for $h=0.5$.}
    \label{fig:comp1}
\end{figure}

Figure \ref{fig:comp1} shows the entries of the density matrix of the pure undeformed Bell state given by $\ket{\Psi^-}= (\ket{\up\down}-\ket{\down\up })/\sqrt{2}$ as well as the corresponding $q$- and $h$-deformed states $\ket{M^0_2}$ for $q=h=0.5$. As usual, the squares on the diagonal (i.e., ~$\rho_{ii}$) represent the populations or occupancy probabilities of each base state. The off-diagonal elements ($\rho_{ij},\,  \textrm{with} \,\, i\neq j$) represent the quantum coherences or interference amplitudes between the states. 

Interestingly, these types of diagrams are experimentally achievable (see Fig.~3 in \cite{Kimbleqinternet}{~\footnote{{According to the DLCZ protocol described in Ref.~\cite{Kimbleqinternet}, the relative sign of the entangled state $\ket{\Psi_{L,R}} = \frac{1}{\sqrt{2}}(\ket{\up\down } \pm e^{i\eta_1}\ket{\down \up})$ depends on the detector that registers the heralding photon. The density matrix in Figure 3b in \cite{Kimbleqinternet} corresponds to the case in which $\eta_1 = 0$ and the positive sign is selected, resulting in constructive interference and positive off-diagonal elements. Our state in Fig. 2 uses the negative sign (singlet state), which correctly produces negative off-diagonal coherences. The weights of each entry in the density matrix of Fig. 3b in \cite{Kimbleqinternet} can then be directly compared with our results of Fig.~\ref{fig:comp1} once we take the absolute value.}}}). The experimental setup involves two spatially separated atomic ensembles, synchronized laser pulses for entanglement generation, photon interference at a beam splitter, heralded entanglement creation, and quantum state tomography~\cite{poyatos1996motion} to verify the entangled states. Fig.~3 in \cite{Kimbleqinternet} presents the density matrix associated with the entangled state stored between two distant atomic ensembles, as inferred from the experimental data. This matrix is quite close to the $q$-Dicke state $\ket{M^0_2}_{q=0.5}$. Moreover, \cite{Stemp24} reported the first experimental demonstration of universal quantum logic operations between electron spins bound to individual phosphorus donors in silicon. The electron spins of the donor atoms serve as qubits, and their quantum states and gate operations are shown in Fig.~5. of \cite{Stemp24}.  The system is controlled and read out using standard semiconductor techniques, including electron spin resonance (ESR) for qubit manipulation and a single-electron transistor (SET) for high-fidelity, single-shot spin readouts. The histograms concerning the measured probabilities of the spin states of the two-electron qubit system found in each computational basis state after a two-qubit gate operation in Fig. 5 of that work are more reminiscent of the non-symmetric configurations obtained with the $h$-deformed states $\ket{M^0_2}_{h}$ which we propose here. The results of \cite{Stemp24} pave the way for large-scale silicon-based quantum computers using donor electron spins. In addition, in~\cite{kiesel2007experimental}, an ultraviolet laser was used to bombard a barium borate crystal, occasionally producing two pairs of photons through the spontaneous parametric down conversion (SPDC)~\cite{krischek2010ultraviolet}  type II process (which means that the photons generated have orthogonal polarizations) and in a collinear configuration (viz.  photons emerge in the same direction). These four photons are used to prepare and study the Dicke state of four particles $\ket{D^2_4}$. Notice that advanced techniques (such as quantum error correction codes and system symmetries) are being used to make the experimental reconstruction of complex quantum states more efficient~\cite{stricker2020experimental,stricker2022characterizing}. Indeed, based on permutationally invariant tomography~\cite{schwemmer2014experimental},  the Dicke states of six spins are experimentally obtained using SPDC techniques. In addition, by applying certain filters or thresholds in the data analysis~\cite{binosi2024tailor}, the number of measurements required can be reduced without losing significant accuracy in the quantum-state description (losses of approximately 1$\%$ only).  In all these papers, the entanglement criteria can be studied in terms of the density matrix elements. However, these are not the only possible applications reported in the literature. In quantum computing applications, Dicke states have been prepared using IBM quantum computers ~\cite{cruz2019efficient,aktar2022divide}. In these papers, density matrices similar to those obtained in this work appear. Therefore, the states proposed in this section and our characterization of them would allow us to visualize this density matrix and compare it with the experimental thresholds of the above-mentioned papers.

\subsection{Schmidt decomposition of \texorpdfstring{$h$}{h}-states}

Here, we present in detail the identification of the $h$-states in Eqs. \eqref{eq:2qubits} and \eqref{Mh} in terms of the Schmidt decomposition \cite{Schmidt1907,ekert1995entangled,acin2000generalized}, which provides a way to represent any pure state of a bipartite quantum system as a sum of product states that are mutually orthogonal, constructed from suitable local bases for each subsystem. This approach ensures that the expansion uses only a minimal set of necessary product states with no redundancy. In order to make it clear, we start with a two-qubit state $\ket{\varphi_\theta}$ in the undeformed scenario, which reads
\begin{equation}
    \ket{\varphi_\theta} = \cos \theta  \ket{\up \up}  + \sin \theta \ket{\down \down} = 
    \begin{pmatrix}
        \cos \theta \\
        0 \\
        0 \\
        \sin \theta \\
    \end{pmatrix}, \qquad\qquad \theta \in \left[0, \frac{\pi}{4}\right] .
    \label{eq:schmidt}
\end{equation}
In this context, $\ket{ii}$ denotes the product state $\ket{i}_A \otimes \ket{i}_B$, where both sets of local basis states, $\{\ket{i}\}_A$ and $\{\ket{i}\}_B$, are chosen specifically for the given two-qubit state $\ket{\varphi_\theta}$. Any relative phase can be incorporated into either local basis. By convention, the state $\ket{\up \up}$ is associated with the largest Schmidt coefficient. A greater value of parameter $\theta$ corresponds to a higher degree of entanglement. Altogether, a pure two-qubit state is characterized by six real parameters: the entanglement parameter $\theta$, a hidden relative phase, and two parameters for each local basis\footnote{Each local basis (that of the first qubit and that of the second qubit) can be specified with two real parameters each, because any orthonormal basis on $\mathbb{C}^2$ (the space of a qubit) can be obtained by a general rotation of the Bloch sphere, which requires two angles to describe a unit vector on the Bloch sphere.}, once the normalization and global phase have been fixed. 

The Schmidt decomposition is important. In large-scale simulations (such as those employing matrix product states, the density matrix renormalization group, or general tensor network techniques), maintaining the Schmidt decomposition at each system bipartition enables one to optimally truncate the less relevant degrees of freedom~\cite{schollwock2011dmrg}. This approach ensures computational efficiency by retaining only the most significant contributions to entanglement spectra. Furthermore, Schmidt form is invaluable for designing optimal local operations. For any protocol that manipulates the quantum state (such as teleportation, entanglement concentration, or conversions between local and classical resources), knowledge of the Schmidt decomposition directly identifies which local unitaries to implement and with what probabilities, thereby enabling precise control over the transformation process \cite{nielsen1999conditions}.

For the non-trivial $h$-deformed state of the triplet in \eqref{eq:2qubits}, we define two local unitary operators $U_1$ (acting on the first qubit) and $U_2$	(acting on the second qubit), whose explicit matrix forms are 
{\begin{equation}
U_1=-\frac{1}{b}\begin{pmatrix}
        4 \, \text{sgn}(h) & -\frac{4+h^2+a}{h} \,  \text{sgn}(h) \\
           \frac{4+h^2+a}{h} & 4
    \end{pmatrix},\qquad  U_2=\frac{1}{c}\begin{pmatrix}
        \frac{a-4-h^2}{h} &  4\\
         -4 \, \text{sgn}(h)& \frac{a-4-h^2}{h} \, \text{sgn}(h)
    \end{pmatrix},
\end{equation}}
where {where sgn($h$)=1 for $h\geq0$ and sgn($h$)=$-1$ for $h<0$, } and
\begin{equation}\label{abccoef}
a=\sqrt{16+24h^2+h^4},\qquad b=\sqrt{16+\frac{(4+h^2+a)^2}{h^2}},\qquad c=\sqrt{16+\frac{(4+h^2-a)^2}{h^2}}.
\end{equation}
These unitary matrices are chosen such that{\footnote{{In the process of selecting matrices for the decomposition of Eq.~\eqref{eq:d22sd}, a degree of arbitrariness exists. To mitigate this, we constrained the elements of these matrices to real numbers. Subsequently, any residual arbitrariness is further minimized by prioritizing computational simplicity.}}}
\begin{equation}
\ket{\tilde D^{-2}_2}_h = U_1 \otimes U_2 \ket{D^{-2}_2}_h, \end{equation}
transforms the original state into its canonical Schmidt form
\begin{equation}
   \ket{\tilde D^{-2}_2}_h ={\frac{1}{2(4+h^2)}\begin{pmatrix}
     a-h^2+4   \\
        0  \\
       0   \\
     a+h^2-4  \\
    \end{pmatrix}} ,
    \label{eq:d22sd}
\end{equation}
Here, the state is written as a column vector with only two non-zero components, which are precisely the Schmidt coefficients.

Similarly, for the $h$-deformed state of the singlet in \eqref{eq:2qubits} we find 
 \begin{equation}
    \ket{\tilde M^{0}_2}_h = U_1 \otimes U_2 \ket{M^{0}_2}_h, \end{equation}
where {
     \begin{equation}
    U_1=\frac{1}{d_-}\begin{pmatrix}
        2 &  -h+\sqrt{4+h^2} \\
            h-\sqrt{4+h^2} & 2
    \end{pmatrix},\qquad  U_2=-\frac{1}{d_+}\begin{pmatrix}
   h+\sqrt{4+h^2}  &  -2    \\
           2   &      h+\sqrt{4+h^2}
    \end{pmatrix},
\end{equation}}
and {
\begin{equation}\label{dcoef}
d_\pm=\sqrt{2}\sqrt{4+h^2\pm h\sqrt{4+h^2}}.
\end{equation}}
Again, we can write the state into its Schmidt decomposition form as {
\begin{equation}
\ket{\tilde M^{0}_2}_h = \frac{1}{\sqrt{h^2+2}}\begin{pmatrix}
      \frac{d_+}{d_-}  \\
        0  \\
       0   \\
       \frac{d_-}{d_+}   \\
    \end{pmatrix}.
\end{equation}}
Up to individual phase factors on each $\ket{i}_A$ and $\ket{i}_B$ the Schmidt decomposition is unique. This simplifies the comparison of states and the classification of state families based on their coefficients. For the deformed states analyzed in this section, the Schmidt decomposition yields exactly two nonzero Schmidt coefficients. Since the Schmidt rank is greater than 1, we know that both deformed states remain entangled (i.e., they are not separable).

\section{3-qubit \texorpdfstring{$h$}{h}-states}
\label{sec:3q}
Let us now consider the space of states resulting from the tensor product of three spin-$1/2$ representations. Using the Clebsch-Gordan coefficients for $\mathcal{U}_h (\mathfrak{sl}(2, \mathbb{R}))$, we obtain the following basis states for the irreducible components of the direct sum decomposition of the tensor product representation:
\begin{eqnarray}
&&\ket{D^{-3}_3}_h =\frac{4}{\sqrt{19h^4+32h^2+16}}\left(\ket{\down\down \down}-h\ket{ \up\down\down}+h\ket{ \down\down\up}+\frac{3h^2}{4}\ket{\up\up\down} -\frac{h^2}{4}\ket{\up \down\up} +\frac{3h^2}{4}\ket{\down\up\up} \right),	\nonumber\\
&&{\ket{D^{-1}_3}_h =	\frac{4 }{\sqrt{9h^4+32h^2+48}}\left(\ket{ \up\down\down} + \ket{ \down\up \down} +\ket{ \down\down\up}  -h\ket{\up\up \down} +h\ket{\down \up\up}+\frac{3h^2}{4}\ket{\up\up\up} \right),}	\nonumber\\
&&\ket{D^1_3}_h =\frac{1}{\sqrt{3}} (\ket{ \up\up\down} + \ket{ \up\down\up } +\ket{ \down\up\up} ),\nonumber\\
&&\ket{D^3_3}_h =|\up\up\up\rangle,
\label{eq:3qubits}
\end{eqnarray}
\begin{eqnarray}
&&{\ket{M^{-1}_3}_h =	\frac{1}{\sqrt{5h^2+6}}\left( \ket{ \up\down\down }  + \ket{ \down\up\down}-2 \ket{ \down\down\up}-h(\ket{ \up\up\down}+2\ket{ \down\up\up})\right)},\nonumber\\
&&{\ket{M^{1}_3}_h=	\frac{1}{\sqrt{3(3h^2+2)}}\left(2\ket{\up\up \down}-\ket{\up\down \up}-\ket{\down\up \up}-3h \ket{\up\up \up}\right)},\\[2.5ex]
&&\ket{V^{-1}_3}_h={\frac{1}{\sqrt{h^2+2}}\left(  \ket{ \up\down\down}-\ket{ \down\up\down}-h \ket{ \up\up\down}	\right)},\nonumber\\
&&\ket{V^{1}_3}_h ={ \frac{1}{\sqrt{h^2+2}}\left(\ket{ \up\down\up}-\ket{ \down\up\up}-h \ket{ \up\up\up}\right)}.
\label{eq:3qubitsa}
\end{eqnarray}
We have used the notation shown in Fig.~\ref{fig:three_spin_tree_DUV} to label these states.
\begin{figure}[h!]
    \centering
    \begin{tikzpicture}[
        level distance=1.5cm,
        level 1/.style={sibling distance=4cm},
        level 2/.style={sibling distance=3cm},
        every node/.style={font=\small, align=center},
        vertical child/.style={
            edge from parent path={(\tikzparentnode.south) -- (\tikzchildnode.north)}
        }
    ]
        \node {$\tfrac{1}{2}\;\oplus\;\tfrac{1}{2}$}
            child {
                node {$1 \oplus\tfrac{1}{2}$}
                child {
                    node {$\tfrac{3}{2}$}
                    child[vertical child] { node {$D$} }
                }
                child {
                    node {$\tfrac{1}{2}$}
                    child[vertical child] { node {$M$} }
                }
            }
            child {
                node {$0\oplus \tfrac{1}{2} $}
                child {
                    node {$\tfrac{1}{2}$}
                    child[vertical child] { node {$V$} }
                }
            }
        ;
    \end{tikzpicture}
    \caption{Coupling tree for three spin-\(\tfrac{1}{2}\) particles, with final labels \(D\), \(M\) and \(V\). }
    \label{fig:three_spin_tree_DUV}
\end{figure}
The quadruplet of $h$-Dicke states (with $j=3/2$) are given by two undeformed ones $\{\ket{D^1_3}_h , \ket{D^3_3}_h\}$, together with the deformed states $\{\ket{D^{-3}_3}_h, \ket{D^{-1}_3}_h\}$, which are symmetric under the exchange $h \to - h$ together with the permutation $\ket{ i\ell k}\to \ket{ k\ell i}$. Furthermore, we again observe that the states $\{\ket{V^{-1}_3}_h, \ket{V^{1}_3}_h\}$ ($j=1/2$) remain unchanged when $\ket{ i\ell k}\to \ket{ \ell ik}$ if we also make $h \to - h$ and add a global sign. As expected, this symmetry did not hold for the other two doublets.

\begin{figure}[H]
    \centering
    \includegraphics[width=1\linewidth]{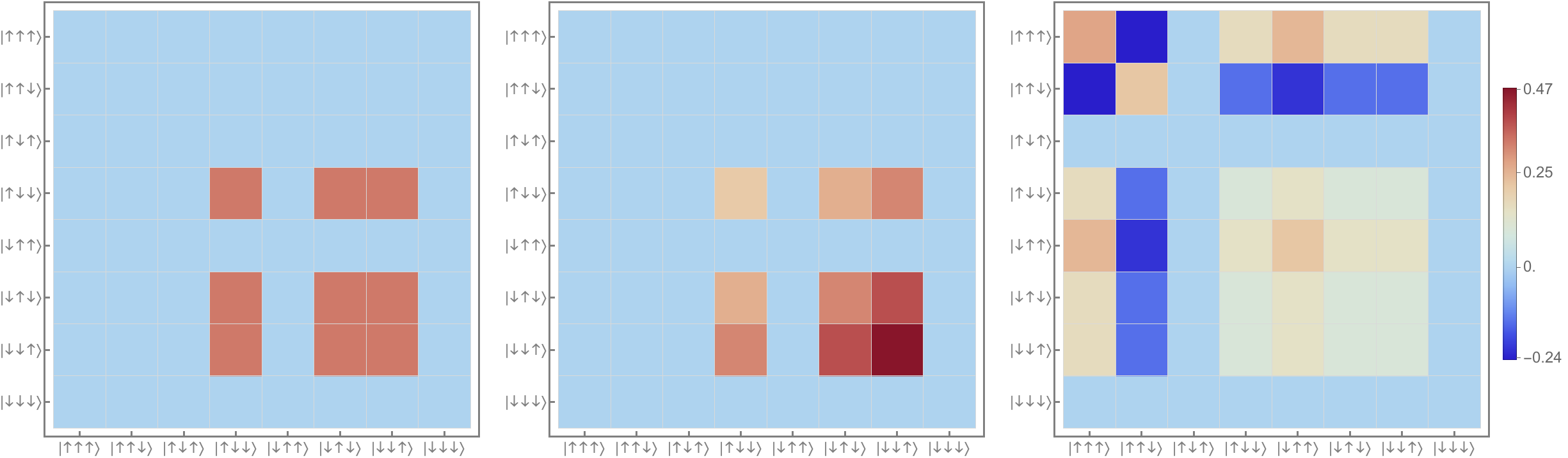}
    \caption{\small Density matrix for the undeformed state $\ket{D^{-1}_3}_{h=0}$ (left) and for the $h$-Dicke deformed one $\ket{D^{-1}_3}_{h=1.5}$ (right). Density matrix for the $q$-Dicke deformed state $\ket{D^{-1/2}_3}_{q=1.5}$ (middle).}
    \label{fig:comp3qubitsh}
\end{figure}

It is worth stressing that, as a general feature of the associated $h$-Clebsch-Gordan coefficients, the $h$-deformation generates a factor $h^k$ in the coefficients for states with $k$ extra excitations with respect to the undeformed state. This also occurs for the 2-qubits states of the previous section, and as shown in Table \ref{table:4q} in Appendix \ref{AppB}, it is also true for 4-qubits. As the value of $h$ increases, the deformed state moves further away from its undeformed analog (see Fig. ~\ref{fig:comp3qubitsh}). It is worth stressing that the experimental results of \cite{cruz2019efficient} resemble our state $\{\ket{D^{-1}_3}_h$ for small $h$.

On the other hand, the quadruplet of $q$-Dicke states (with $j=3/2$) are given by \cite{Ballesteros:2025cia}
\begin{eqnarray}
&&\ket{D^{-3/2}_3}_q =|\hspace{-0.1cm}\down\down\down\rangle,	\nonumber\\
&&\ket{D^{-1/2}_3}_q =	\frac{1}{\sqrt{[3]_q}}\left( q^{1/2}|\hspace{-0.1cm}\down\down\up\rangle +|\hspace{-0.1cm}\down\up\down\rangle+ q^{-1/2}|\hspace{-0.1cm}\up\down\down\rangle\right),	\nonumber\\
&&\ket{D^{1/2}_3}_q =\frac{1}{\sqrt{[3]_q}}\left( q^{1/2} |\hspace{-0.1cm}\down\up\up\rangle +|\hspace{-0.1cm}\up\down\up\rangle + q^{-1/2}|\hspace{-0.1cm}\up\up\down\rangle \right),\nonumber\\
&&\ket{D^{3/2}_3}_q =|\up\up\up\rangle,
\label{eq:3qubitsq}
\end{eqnarray}
In Fig.~\ref{fig:comp3qubitsh} we can again see that the $q$-deformed states are such that the weights of each component of the corresponding density matrix change monotonically without adding any extra basis vectors with respect to the undeformed state, while this is no longer the case in the non-standard deformation.

\subsection{Acín classification}
The so called Ac\'{i}n decomposition~\cite{acin2000generalized} allows the equivalence of any tripartite spin-$1/2$ system with a state of the form
\begin{equation}\label{Acin}
    \ket{\phi}_{\lambda_i,\varphi} = \lambda_0 \ket{\up \up \up} + \lambda_1 e^{i \varphi} \ket{\down \up \up} + \lambda_2 \ket{\down \up \down} + \lambda_3 \ket{\down \down \up} + \lambda_4 \ket{\down \down \down},
\end{equation}
where $\lambda_i \ge 0$, $0 \le \varphi \le \pi$ and $\sum_{i=0}^4 \lambda_i^2 = 1$. 

Thus, the objective is to define three local unitary operators $U_1$ (acting on the first qubit), $U_2$	(acting on the second qubit), and $U_3$ (acting on the third qubit) such that $ U_1 \otimes U_2 \otimes U_3$ transforms the original $h$-deformed state into its canonical Acin form \eqref{Acin}. The explicit expressions for these unitary transformations are listed in Tables \ref{table1:acin} and \ref{table2:acin} in Appendix \ref{ap1}.


\section{Towards generic \texorpdfstring{$h$}{h}-Dicke states}
\label{sec:gen}
The aim of this section is to develop some algebraic tools in order to generate $h$-Dicke states explicitly, since a generic closed formula cannot be directly obtained, in contrast to the $q$-Dicke states.  We begin by obtaining a simple form of the operator $\Delta_h^{(N)}(Z_+)$. Then, we provide a simple form for the $\ket{D^{-N}_N}_h$ states, so by applying $\Delta_h^{(N)}(Z_+)$ to the lowest-weight state, one can easily obtain any $h$-Dicke state. Finally, as a relevant example for applications, we provide the explicit construction of the $\ket{W}_h$ states.

\subsection{\texorpdfstring{$\Delta_h^{(N)}(Z_+)$}{dz} for spin-1/2 particles}

The Clebsch--Gordan coefficients for larger chains of qubits can be calculated in an analogous manner by following the procedure explained in the first section of this article (Eqs. \eqref{boots1}, \eqref{boots2}, \eqref{program1}, \eqref{program2}).  For instance, Table \ref{table:4q} in Appendix \ref{AppB} shows all the $h$-deformed states for $4$ qubits as a linear combination of the elements of the initial basis (in the first row of the table) computed in terms of the associated Clebsch--Gordan coefficients. The  notation for such states is shown in Figure~\ref{fig:recoupling_vertical}.

\begin{figure}[h!]
    \centering
    \begin{tikzpicture}[
        level distance=1.4cm,                              
        level 1/.style={sibling distance=5cm},            
        level 2/.style={sibling distance=3cm},            
        level 3/.style={sibling distance=2cm},            
        every node/.style={font=\small, align=center},    
        vertical child/.style={
            edge from parent path={(\tikzparentnode.south) -- (\tikzchildnode.north)}
        }
    ]
        \node {$\tfrac{1}{2}\;\oplus\;\tfrac{1}{2}$}
            child {
                node {$1\;\oplus\;\tfrac{1}{2}$}
                child {
                    node {$\tfrac{3}{2}\;\oplus\;\tfrac{1}{2}$}
                    child {
                        node {$2$} 
                        child[vertical child] { node {$D$} }
                    }
                    child {
                        node {$1$}
                        child[vertical child] { node {$M$} }
                    }
                }
                child {
                    node {$\tfrac{1}{2}\;\oplus\;\tfrac{1}{2}$}
                    child {
                        node {$1$} 
                        child[vertical child] { node {$V$} }
                    }
                    child {
                        node {$0$}
                        child[vertical child] { node {$T$} }
                    }
                }
            }
            child {
                node {$0 \oplus \tfrac{1}{2}$}
                child {
                    node {$\tfrac{1}{2}\oplus \tfrac{1}{2}$}
                    child {
                        node {$1$}
                        child[vertical child] { node {$R$} }
                    }
                    child {
                        node {$0$}
                        child[vertical child] { node {$Q$} }
                }
            }}
        ;
    \end{tikzpicture}
    \caption{Four spin-1/2 coupling tree with vertical lines pointing to state labels.}
    \label{fig:recoupling_vertical}
\end{figure}

However, instead of considering the aforementioned procedure involving the repeated application of the Clebsch--Gordan coefficients, we stress that there exists another, more direct way to obtain the $h$-Dicke states for $N$ qubits.

Generally, when composing spins in quantum mechanics, one calculates the state with the largest eigenvalue of $J_z$ and (maximum  $J$ eigenvalue for the case of Dicke states) and then applies the coproduct of the lowering operator to obtain all the remaining states within the irreducible representation. Equivalently, one can start with the state with the smallest eigenvalue of $J_z$ and the maximum $J$ eigenvalue and apply the coproduct of the raising operator. Both approaches are straightforward to apply to undeformed and $q$-Dicke states. Indeed, it is possible to write a simple formula to obtain the $q$-Dicke states~\cite{li2015entanglement}, since $q$-deformed coproducts~\eqref{qcop} are quite simple. 

Nevertheless, the coproducts in the $h$-deformation~\eqref{eq:coproduct_j} are not so simple, making it difficult to obtain compact analytical general formulas for $h$-Dicke states with any number of qubits. Moreover, in the $h$-deformed case note that, on the one hand, the coproduct of $Z_-$ is much more complicated than the one for $Z_+$ (see \eqref{eq:coproduct_j}).  On the other hand, $\ket{D^{-N}_N}$ is not simply the state with all spins down (see Eqs.~\eqref{2h_dicke} and~\eqref{eq:3qubits}), as is the case in the $q$-deformation, because $h$-contributions arise. We have two possible ways to address the aforementioned difficulties: either by applying the complicated coproduct given by $[\Delta_h^{(N)}(Z_-)]^i$ to a simple known state $(\ket{D_N^{N}}_h= \ket{\up \dots \up})$ in order to obtain the $(\ket{D_N^{N-i}}_h$ state, or by applying a much simpler operator $[\Delta_h^{(N)}(Z_+)]^i$ to a (yet unknown) state $(\ket{D_N^{-N}}_h)$ to obtain the $(\ket{D_N^{-N+i}}_h$ state. Here, we will follow the second option, which will be shown to be feasible. Indeed, within this path the first step is then the computation  of the $h$-deformed Dicke state characterized by the smallest eigenvalue of $H=2J_z$ and maximum $J$ eigenvalue.\footnote{ Notice that the procedure we will explain in this section is only valid for Dicke states $\ket{D}_h$, i.e.~the ones with maximum value of $J$ for a given composition of $N$ spins.} However, first, we must prove one result concerning $\Delta_h^{(N)}(Z_+)$ which will be crucial in the following.

The expression for $\Delta_h(Z_+)$ given in Eq.~\eqref{eq:coproduct_j} can be simplified for the particular representation we are considering, that is, any state $\ket{\Psi}$ resulting from the tensor product of individual spins with angular momentum equal to 1/2, since $	Z_+^{2}\ket{\down}=0$. Indeed, from~\eqref{eq:coproduct_j} and by applying~\eqref{fla}, the expression for the $N$-th $h$-deformed coproduct of $Z_+$ acting on a sate $\ket{\Psi}$ belonging to $[\mathbb{C}^{(2)}]^{\otimes (N)}$ reads 
\begin{align}
\Delta_h^{(N)}(Z_+)\ket{\Psi} &= \sum_{i=1}^N \left( 1^{\otimes (i-1)} \otimes Z_+ \otimes 1^{\otimes (N-i)} \right)\ket{\Psi} \nonumber\\
&\quad - \frac{h^2}{2} \sum_{1 \le i < j < k \le N} \left( 1^{\otimes (i-1)} \otimes Z_+ \otimes 1^{\otimes (j-i-1)} \otimes Z_+ \otimes 1^{\otimes (k-j-1)} \otimes Z_+ \otimes 1^{\otimes (N-k)} \right)\ket{\Psi}.
\label{eq:z+simp}
\end{align}
This can be proven by induction as follows. For $N=2$, expression~\eqref{eq:z+simp} is straightforwardly checked. Then, assuming that the above expression is true for $(N-1)$ and $\ket{\Psi}\in[\mathbb{C}^{(2)}]^{\otimes (N-1)}$ particles, and by taking into account that 
	\begin{equation}
 \Delta_h^{(N)}=(\Delta_h^{(N-1)}\otimes 1)\circ \Delta_h^{(2)} \, ,
\end{equation}
holds, where $\Delta_h^{(2)}$ is given by Eq.~\eqref{eq:coproduct_j}, we obtain
\begin{align}
	\Delta_h^{(N)}(Z_+)\ket{\Psi} &=\sum_{i=1}^{N-2} \left( 1^{\otimes (i-1)} \otimes Z_+ \otimes 1^{\otimes (N-i)} \right)\ket{\Psi}\nonumber\\
    &+1^{\otimes (N-2)}\otimes(1\otimes Z_+ + Z_+ \otimes 1)\left(\sum_{n=0}^\infty \left(-\frac{h^2}{4}\right)^n
 \; Z_+^n \otimes Z_+^n \right) \ket{\Psi}\nonumber\\
  &- \frac{h^2}{2} \sum_{1 \le i < j < k \le N-2} \left( 1^{\otimes (i-1)} \otimes Z_+ \otimes 1^{\otimes (j-i-1)} \otimes Z_+ \otimes 1^{\otimes (k-j-1)} \otimes Z_+ \otimes 1^{\otimes (N-k)} \right)\ket{\Psi}\nonumber\\
  &-  \frac{h^2}{2} \sum_{1 \le i < j \le N-2} \left( 1^{\otimes (i-1)} \otimes Z_+ \otimes 1^{\otimes (j-i-1)} \otimes Z_+  \otimes 1^{\otimes (N-j-2)} \right)\nonumber\\
  \quad&\otimes(1\otimes Z_+ + Z_+ \otimes 1)\left(\sum_{n=0}^\infty \left(-\frac{h^2}{4}\right)^n
 \; Z_+^n \otimes Z_+^n \right)\ket{\Psi},
\end{align}
which automatically leads to Eq.~\eqref{eq:z+simp} when the previous expression is applied onto any state $\ket{\Psi}\in [\mathbb{C}^{(2)}]^{\otimes (N)}$  and taking into account that $Z_+^{n}\ket{\down}=0$ and $Z_+^{n}\ket{\up}=0$ for $n\geq 2$.

Consequently, one can construct all $h$-Dicke states by repeatedly applying the coproduct~\eqref{eq:z+simp} to $|D^{-N}_N\rangle_h$. In fact, the result~\eqref{eq:z+simp} tells us that any $h$-Dicke state $|D^{-N+2i+2}_N\rangle_h$ can be obtained from $|D^{-N+2i}_N\rangle_h$ by summing all its possible flips on one spin ($\down$ to $\up$) plus all its possible flips on three spins multiplied by  $-h^2/2$. After normalization of the states, one can verify that for all the cases shown here ($N=2,3,4$), this formula is indeed valid. The main task now is to obtain a generic formula for the $|D^{-N}_N\rangle_h$ states. This is addressed in the following subsection.

\subsection{Construction of \texorpdfstring{$|D^{-N}_N\rangle_h$}{d}}
The quantum state of interest is a superposition of computational basis states $\ket{S} = \ket{s_1 s_2 \dots s_N}$ (where $s_k \in \{\up, \down\}$ for $1\leq k\leq N$)
\begin{equation}\label{eq58}
    |D^{-N}_N\rangle_h = \sum_{S} \chi_{N}(S) |S\rangle.
\end{equation}
The coefficients $\chi_{N}(S)$ have a particular structure related to the number of $\up$ states ($n_\up$) in the state $|S\rangle$. Let $P_\up(S) = \{k | s_k = \up\}$ be the set of indices where state $S$ is '$\up$', and $n_\up(S) = |P_\up(S)|$ (the cardinal of the set). Looking at the examples for up to 4-qubits states, we observe that the structure of the coefficient is
\begin{equation}\label{eq59}
    \chi_{N}(S) = {\xi}(S) \left(\frac{h}{2}\right)^{n_\up(S)},
\end{equation}
where ${\xi}(S)$ is a factor (generally integer) that depends on $N$ and the specific configuration of $\up$ and $\down$ states in $S$ (i.e., on the set $P_\up(S)$). 

We have discovered that ${\xi}(S)$ can be elegantly expressed in terms of the elementary symmetric polynomials.
We define the function $f(k)$ for an index $k$:
\begin{equation}
    f(k) = 2k - N - 1.
\end{equation}
For a state $|S\rangle$ with $n_\up$  states $\ket{\up}$ at positions $P_\up(S)$, we define $e_m(P_\up(S))$ as the $m$-th elementary symmetric polynomial of the values $\{f(k) | k \in P_\up(S)\}$:
\begin{align}
    e_0(P_\up) &= 1,\nonumber \\
    e_1(P_\up) &= \sum_{k_1 \in P_\up} f(k_1), \nonumber\\
    e_2(P_\up) &= \sum_{k_1 < k_2 \in P_\up} f(k_1)f(k_2), \nonumber\\
    &\dots \nonumber\\
    e_m(P_\up) &= \sum_{k_1 < k_2 < \dots < k_m \in P_\up} f(k_1)f(k_2)\dots f(k_m).
\end{align}
By convention, $e_m(P_\up) = 0$ if $m > n_\up$.

The factor ${\xi}(S)$ for a state with $n_\up$ states $\ket{\up}$ at positions $P_\up$ is given by the sum
\begin{equation}
    {\xi}_{n_\up}(P_\up) = \sum_{m=0}^{\lfloor n_\up/2 \rfloor} C_m(N) e_{n_\up-2m}(P_\up),
    \label{eq:gamma_general}
\end{equation}
where $C_m(N)$ are polynomial coefficients in $N$.

We have determined the first six coefficients $C_m(N)$ by determining the $\ket{D^{-N}_N}$ states for $N=2$ up to $N=11$, which read:
\begin{align}
    C_0(N) &= 1, \nonumber\\
    C_1(N) &= N, \nonumber\\
    C_2(N) &= N(3N-2), \nonumber\\
    C_3(N) &= N(15N^2 - 30N + 16), \nonumber\\
    C_4(N) &= N(105N^3 - 420N^2 + 588N - 272), \nonumber\\
    C_5(N) &= N(945N^4 - 6300N^3 + 16380N^2 - 18960N + 7936).
\end{align}
The leading term of $C_m(N)$ is $(2m-1)!! N^m$, where $(2m-1)!!$ is the odd double factorial.

Using the general formula \eqref{eq:gamma_general} and the known coefficients $C_m(N)$ we can now compute ${\xi}_{n_\up}$, for instance:
\begin{itemize}
    \item ${\xi}_0 = C_0 e_0 = 1$,
    \item ${\xi}_1 = C_0 e_1 = e_1$,
    \item ${\xi}_2 = C_0 e_2 + C_1 e_0 = e_2 + N$,
    \item ${\xi}_3 = C_0 e_3 + C_1 e_1 = e_3 + N e_1$,
    \item ${\xi}_4 = C_0 e_4 + C_1 e_2 + C_2 e_0 = e_4 + N e_2 + N(3N-2)$,
    \item ${\xi}_5 = C_0 e_5 + C_1 e_3 + C_2 e_1 = e_5 + N e_3 + N(3N-2) e_1$,
    \item ${\xi}_6 = C_0 e_6 + C_1 e_4 + C_2 e_2 + C_3 e_0 = e_6 + N e_4 + N(3N-2) e_2 + N(15N^2 - 30N + 16)$,
    \item ${\xi}_7 = C_0 e_7 + C_1 e_5 + C_2 e_3 + C_3 e_1 = e_7 + N e_5 + N(3N-2) e_3 + N(15N^2 - 30N + 16) e_1$,
    \item ${\xi}_8 = C_0 e_8 + C_1 e_6 + C_2 e_4 + C_3 e_2 + C_4 e_0 = e_8 + N e_6 + N(3N-2) e_4 + N(15N^2 - 30N + 16) e_2 + N(105N^3 - 420N^2 + 588N - 272)$,
    \item ${\xi}_9 = C_0 e_9 + C_1 e_7 + C_2 e_5 + C_3 e_3 + C_4 e_1 = e_9 + N e_7 + N(3N-2) e_5 + N(15N^2 - 30N + 16) e_3 + N(105N^3 - 420N^2 + 588N - 272) e_1$,
    \item ${\xi}_{10} = C_0 e_{10} + C_1 e_8 + C_2 e_6 + C_3 e_4 + C_4 e_2 + C_5 e_0 = e_{10} + N e_8 + N(3N-2) e_6 + N(15N^2 - 30N + 16) e_4 + N(105N^3 - 420N^2 + 588N - 272) e_2 + N(945N^4 - 6300N^3 + 16380N^2 - 18960N + 7936)$,
    \item ${\xi}_{11} = C_0 e_{11} + C_1 e_9 + C_2 e_7 + C_3 e_5 + C_4 e_3 + C_5 e_1 = e_{11} + N e_9 + N(3N-2) e_7 + N(15N^2 - 30N + 16) e_5 + N(105N^3 - 420N^2 + 588N - 272) e_3 + N(945N^4 - 6300N^3 + 16380N^2 - 18960N + 7936) e_1$.
\end{itemize}

\subsection{Generic \texorpdfstring{$h$}{h}-deformed \texorpdfstring{$W$}{W}-states}
From the results of the two previous subsections, we can construct any $h$-Dicke state $\ket{D_N^{-N+i}}_h$ by applying the operator $[\Delta_h^{(N)}(Z_+)]^i$ to the state $\ket{D_N^{-N}}_h$. In this subsection, we provide an expression for the  $\ket{W_N}_h$ states, which are the $h$-deformed analogs of the $\ket{D_N^{-N+1}}$ states. These $h$-deformed states generally have basis states with $(k+1)$ excitations $\ket{\up}$ multiplied by a factor containing $h^k$. For $N=2$ and 3, they are given by the second equation of~\eqref{2h_dicke} and \eqref{eq:3qubits}, respectively, and for $N=4$ they can be written straightforwardly from Table \ref{table:4q} of Appendix \ref{AppB}. For completion, we give its explicit form, which reads
\begin{align} \label{Wh4}
	\ket{W_4}_h&=\frac{1}{2}\left(\ket{\uparrow\downarrow\downarrow\downarrow} +  \ket{\downarrow\uparrow\downarrow\downarrow} +  \ket{\downarrow\downarrow\uparrow\downarrow}+\ket{\downarrow\downarrow\downarrow\uparrow}\right)+\frac{h}{2}\left(-2 \ket{\uparrow\uparrow\downarrow\downarrow} - \ket{\uparrow\downarrow\uparrow\downarrow} + \ket{\downarrow\uparrow\downarrow\uparrow} + 2\ket{\downarrow\downarrow\uparrow\uparrow}\right) \\ \nonumber
    & + \frac{1}{8}h^2\left(\ket{\uparrow\uparrow\downarrow\uparrow} + \ket{\uparrow\downarrow\uparrow\uparrow} +9\ket{\uparrow\uparrow\uparrow\downarrow} + 9\ket{\downarrow\uparrow\uparrow\uparrow}\right).
\end{align}


Now, by computing $\Delta_h^{(N)}(Z_+) \ket{D_N^{-N}}_h$ (from Eqs. \eqref{eq:z+simp}, \eqref{eq58}, \eqref{eq59}, \eqref{eq:gamma_general}), one finds 
\begin{equation}\label{eq64}
    |W\rangle_h = \sum_{S'} {\xi}^\prime(S')\left(\frac{h}{2}\right)^{n_\up(S)-1} |S'\rangle,
\end{equation}
where
\begin{align}\label{eq65}
{\xi}^\prime(S') &=  \sum_{m=0}^{\lfloor (\nought{S}-1)/2 \rfloor} (2m+1) \Ccoeff{m} \ePoly{\nought{S}-1-2m}{\pos{S}}  \nonumber\\
& - 2 \sum_{m=0}^{\lfloor (\nought{S}-3)/2 \rfloor} \binom{2m+3}{3} \Ccoeff{m} \ePoly{\nought{S}-3-2m}{\pos{S}}.
\end{align}
After normalization, one finds the $\ket{W}_h$ states shown in this work. 

Let us explain how Eqs.~\eqref{eq64} and \eqref{eq65} are obtained. Firstly, there is a relationship we will use in the following, regarding the elementary symmetric polynomials, 
\begin{equation}
    \sum_{J \subset P_\up(S), |J|=r} e_i(P_\up(S) \setminus J) = \binom{n_\up(S)-i}{r} e_i(P_\up(S)).
    \label{eq:main_identity}
\end{equation}
where the summation runs over all possible subsets $J$ of $r$ elements of $P_\up(S) = \{k | s_k = \up\}$, and $\{P_\up(S) \setminus J\}$ means the set of elements of $P_\up(S)$ that are not in $J$. Notice that $n_\up(S)$ is the cardinal of $P_\up(S)$. This equation, based on the recurrence relation for the elementary symmetric polynomials \cite{ELMIKKAWY20138770, Macdonaldbook}, is proven in Appendix~\ref{appendix_proof}.

Now, the first term of the coproduct \eqref{eq:z+simp}, which leads to the first term in \eqref{eq65}, involves the sum over the action of all possible flips of one $ \down$ qubit to an $\up$ qubit in the initial state $|D^{-N}_N\rangle_h$.  Therefore, since we consider 1-flips, we note that we must replace $\nought{S}$ with $\nought{S}-1$. Moreover, the result of performing all possible 1-flips over $|D^{-N}_N\rangle_h$ yields a new superposition of computational states $S'$ with coefficients given by $(h/2)^{n_\up(S)-1} $ multiplied by
\begin{align}
	&\qquad\sum_{j \in \pos{S}} \sum_{m=0}^{\lfloor (\nought{S}-1)/2 \rfloor}   \Ccoeff{m} \ePoly{\nought{S}-1-2m}{\pos{S} \setminus \{j\}}\nonumber\\ &= \sum_{m=0}^{\lfloor (\nought{S}-1)/2 \rfloor}   \Ccoeff{m}(\nought{S} - (\nought{S}-1-2m)) \ePoly{\nought{S}-1-2m}{\pos{S}}\nonumber \\
   & =\sum_{m=0}^{\lfloor (\nought{S}-1)/2 \rfloor}  (2m+1) \Ccoeff{m}  \ePoly{\nought{S}-1-2m}{\pos{S}}
\end{align}


The second term of the coproduct \eqref{eq:z+simp}, which leads to the second term in \eqref{eq65}, is the sum over all possible flips of three qubits in the starting state $|D^{-N}_N\rangle_h$. First, since we consider 3-flips, we note that we must replace $\nought{S}$ with $\nought{S}-3$. In addition, we need to consider the sum over all unique combinations of three elements $\{j, k, l\}$ removed from the set $\pos{S}$. Therefore, taking into account eq. \eqref{eq:main_identity} for $r=3$, we can write
\begin{align}
    &\qquad\sum_{j<k<l \in \pos{S}} \sum_{m=0}^{\lfloor (\nought{S}-3)/2 \rfloor} \Ccoeff{m} \ePoly{\nought{S}-2m-3}{\pos{S} \setminus \{j,k,l\}}\nonumber \\
    &= \sum_{m=0}^{\lfloor (\nought{S}-3)/2 \rfloor} \Ccoeff{m} \binom{\nought{S} - (\nought{S}-3-2m)}{3} \ePoly{\nought{S}-2m-3}{\pos{S}} \nonumber\\
    &= \sum_{m=0}^{\lfloor (\nought{S}-3)/2 \rfloor} \binom{2m+3}{3} \Ccoeff{m} \ePoly{\nought{S}-2m-3}{\pos{S}}.
\end{align}
The sum vanishes if $\nought{S'} < 3$, as no 3-flip parents can exist. 

Notice that the overall prefactor $(-2)$ in the second term of \eqref{eq65} arises because when applying the second summation of \eqref{eq:z+simp} to $|D^{-N}_N\rangle_h = \sum_{S} {\xi}_{n_\uparrow}(S) \left(\frac{h}{2}\right)^{n_\uparrow(S)} |S\rangle$, we have to change $(h/2)^{\nought{S}}$ by $(h/2)^{\nought{S}-3}$. Thus, the $-h^2/2$ factor in the coproduct combines with the $(2/h)^2$ factor coming from  
\begin{equation}
	 \left(\frac{h}{2}\right)^{\nought{S}-3} = \left(\frac{h}{2}\right)^{\nought{S}-1} \left(\frac{2}{h}\right)^2,
\end{equation}
given this overall prefactor $(-2)$.

\subsection{Considerations for large \texorpdfstring{$N$}{n}}
For a fixed number of $\up$ states $n_\up$ and a very large $N$, there are some caveats that are worth stressing:
\begin{itemize}
    \item As the position $k$ at which there is a qubit $\up$ increases, and if it does so proportionally to $N$ (i.e. $xN$), then the polynomial $f(k) \approx N(2x-1)$, the elementary symmetric polynomials $e_m(P_\up)$ scales as $N^m$, and $C_m(N)$ scales as $(2m-1)!! N^m$. Consequently, in the summation ${\xi}_{n_\up} = \sum_{m=0}^{\lfloor n_\up/2 \rfloor} C_m(N) e_{n_\up-2m}$, the $m$-th term scales as $N^m \times N^{n_\up-2m} = N^{n_\up-m}$.
    \item The dominant term for $N \to \infty$ is the $m=0$ term, i.e. the highest power term of $N$, which is $e_{n_\up}$. This is nothing but the product of $f(k)$ for all positions where there is an $\up$ qubit. Therefore, in this asymptotic limit: ${\xi}_{n_\up}(P_\up) \approx e_{n_\up}(P_\up) = \prod_{k \in P_\up} (2k-N-1)$. The terms with $m\geq 1$ are corrections that become less relevant as $N$ becomes much larger than $n$.
\end{itemize}

Therefore, for a large number of spins, the $|D^{-N}_N\rangle_h$ states can be written as
\begin{equation}
    |D^{-N}_N\rangle_h = \sum_{S} {\xi}_{n_\uparrow}(S) \left(\frac{h}{2}\right)^{n_\uparrow(S)} |S\rangle\approx \sum_{S} \sum_{m=0}^{\lfloor n_\up/2 \rfloor} (2m-1)!! N^m e_{n_\up-2m}(P_\up)\left(\frac{h}{2}\right)^{n_\uparrow(S)} |S\rangle.
\end{equation}
Note that, in this case, we can provide an approximate expression for any number of qubits because we are considering only the dominant term in $C_m(N)$ and not their exact form, which is much more complicated to compute. This could be especially useful in quantum systems with many qubits, where the smallest powers in $N$ of $C_m(N)$ are negligible, and the principal behavior of $N^m$ dominates.

\section{Conclusions}
\label{sec:conclusions}
In this work, we explored the Jordanian quantum algebra $\mathcal{U}_h(sl(2,\mathbb{R}))$ as an algebraic tool that enables us to obtain new quantum states with an arbitrary number of qubits and endowed with a deformation parameter $h$ that can be used to control relevant quantum properties.
We developed a systematic approach to construct $h$-deformed Dicke states for multi-qubit systems, providing explicit expressions for up to four qubits and a general method to construct these states for larger numbers of qubits. As an outstanding example, the construction of generic $\ket{W}_h$ states for any number of qubits is presented in detail.

In particular, we have found that $h$-deformed Dicke states 
are smooth $h$-deformations of the usual Dicke states that include qubit states with $k$ extra excitations endowed with weights of the type  $h^k$. For instance, in a $\ket{W}_h$ state, the first-order deformation in $h$ will include qubit states with two excitations instead of one, which could be considered an algebraic way of describing imperfections in the experimental preparation of a true $\ket{W}$ state (see~\eqref{Wh4}). In the same manner, $h^2$ terms would contain two extra excitations, which should be less likely to occur, and the probability of existence of a state with $k$ extra excitations is ruled by a factor $h^k$. In this sense, our preliminary analysis of the density matrices for $h$-deformed states suggests that these states might be useful in describing noise and decoherence effects in experimental settings. 

Moreover, the $h$-deformation provides a new tool for the generation and study of a novel class of entangled states, whose applications in quantum computing and communication seem worth analyzing in more depth. Also, a detailed study of the $h$-deformation of the rich permutational symmetries of Dicke states and their algebraic structures is worth facing. Additionally, further comparative studies with the $q$-states arising from the standard quantum deformation could provide a more comprehensive understanding of the role of quantum deformations in quantum state characterization and properties. Work on all these aspects is currently in progress.


\section*{Acknowledgments}
The authors acknowledge partial support from the grant PID2023-148373NB-I00 funded by MCIN/ AEI / 10.13039/501100011033/FEDER -- UE, and the Q-CAYLE Project funded by the Regional Government of Castilla y León (Junta de Castilla y León) and the Ministry of Science and Innovation MICIN through NextGenerationEU (PRTR C17.I1).

\bibliography{biblio-3}

\appendix
\section{Ac\'{i}n decomposition for 3-qubit \texorpdfstring{$h$}{h}-deformed states}\label{ap1}
\renewcommand{\arraystretch}{1.8}
\begin{table}[H]
    \centering
   \resizebox{0.6\textwidth}{0.8\textheight}{\Rotatebox{90}{ \begin{tabular}{|c||c|c|c|c|}
       \hline
       
            & { $\ket{V^1_3}_h$}
            & { $\ket{V^{-1}_3}_h$}
            &  {$\ket{M^1_3}_h$}
            &{ $\ket{M^{-1}_3}_h$}
        
        \\ 
        \hline\hline
        $U_1$ 
            & $
        {\frac{1}{d_-}\begin{pmatrix}
        2 &  -h+\sqrt{4+h^2} \\
            h-\sqrt{4+h^2} & 2
    \end{pmatrix}} $
            & $ {\frac{1}{d_-}\begin{pmatrix}
        2 &  -h+\sqrt{4+h^2} \\
            h-\sqrt{4+h^2} & 2
    \end{pmatrix}} $ 
            &{ $\begin{pmatrix}
   0 &\text{ sgn}(h) \\
        1 & 0
    \end{pmatrix}$ }
            &{ $\left(
\begin{array}{cc}
 \frac{1}{\sqrt{4 h^2+1}} & \frac{2 h}{\sqrt{4 h^2+1}} \\
 \frac{2 |h|}{\sqrt{4 h^2+1} \text{ sgn}\left(-4 h^4+3 h^2+1\right)} & -\frac{\text{ sgn}\left(h\right)}{\sqrt{4 h^2+1} \text{ sgn}\left(-4 h^4+3 h^2+1\right)} \\
\end{array}
\right)$ }

        \\ \hline
        $U_2$ 
            &{$
       -\frac{1}{d_+}\begin{pmatrix}
   h+\sqrt{4+h^2}  &  -2    \\
           2   &      h+\sqrt{4+h^2}
    \end{pmatrix} $}
            & {$
       -\frac{1}{d_+}\begin{pmatrix}
   h+\sqrt{4+h^2}  &  -2    \\
           2   &      h+\sqrt{4+h^2}
    \end{pmatrix} $} 
            &{$\begin{pmatrix}
       -1 &0 \\
         0 & -\text{ sgn}(h)
    \end{pmatrix}$} 
            & {$
      \left(
\begin{array}{cc}
 \frac{h}{\sqrt{h^2+1}} & \frac{1}{\sqrt{h^2+1}} \\
 -\frac{\text{ sgn}\left(-4 h^4+3 h^2+1\right)}{\sqrt{h^2+1} \text{ sgn}\left( h\right)} & \frac{|h| \text{ sgn}\left(-4 h^4+3 h^2+1\right)}{\sqrt{h^2+1} } \\
\end{array}
\right)
    $ }
          
        \\ \hline
        $U_3$ 
            &$ \begin{pmatrix}
        1 &0 \\
         0 & 1
    \end{pmatrix}$ 
            & $\begin{pmatrix}
        0 &1 \\
         1 & 0
    \end{pmatrix}$ 
            & {$\begin{pmatrix}
        \text{sign}(h)&0 \\
         0 & -1 
    \end{pmatrix}$ }
            & {$\left(
\begin{array}{cc}
 -\frac{4 h}{\sqrt{16 h^2+1}} & \frac{1}{\sqrt{16 h^2+1}} \\
 \frac{1}{\sqrt{16 h^2+1} \text{ sgn}\left(h\right)} & \frac{4 |h|}{\sqrt{16 h^2+1} } \\
\end{array}
\right)$} 
            
        \\ \hline
        $\lambda_0$ 
            &{$\frac{1}{\sqrt{h^2+2}}\frac{d_+}{d_-}$} 
            & {$\frac{1}{\sqrt{h^2+2}}\frac{d_+}{d_-}$}  
            & {{$\frac{1}{\sqrt{9 h^2+6}}$} }
            &{$\sqrt{\frac{\left(h^2+1\right) \left(16 h^2+1\right)}{20 h^4+29 h^2+6}}$}
            
        \\ \hline
         $\lambda_1$ 
            & 0 
            & 0 
            &{{$\frac{3 |h|}{\sqrt{9 h^2+6}}$} }
            & {$\frac{|h| \left(10 h^2+7\right) }{\sqrt{\left(h^2+1\right) \left(4 h^2+1\right) \left(5 h^2+6\right) \left(16 h^2+1\right)} }$} 
            
        \\ \hline
         $\lambda_2$ 
            & 0 
            & 0
            & {$\frac{2}{\sqrt{9 h^2+6}}$} 
            & {$\frac{\left| -8 h^4+6 h^2+2\right| }{\sqrt{\left(h^2+1\right) \left(4 h^2+1\right) \left(5 h^2+6\right) \left(16 h^2+1\right)}}$}  

        \\ \hline
         $\lambda_3$ 
            & {$\frac{1}{\sqrt{h^2+2}}\frac{d_-}{d_+}$}  
            & {$\frac{1}{\sqrt{h^2+2}}\frac{d_-}{d_+}$}  
            &  {$\frac{1}{\sqrt{9 h^2+6}}$}  
            & {$\sqrt{\frac{4 h^2+1}{\left(h^2+1\right)  \left(5 h^2+6\right) \left(16 h^2+1\right)}}$} 
        \\ \hline
         $\lambda_4$ 
            & 0 
            & 0 
            & 0 
            & {$4 |h| \sqrt{\frac{4 h^2+1}{\left(h^2+1\right)  \left(5 h^2+6\right) \left(16 h^2+1\right)}}$} 
        \\ \hline
         $\varphi$ & 0&0&0& {$-i  \log_{\pi} \left(-\text{ sgn}(1+3h^2-4h^4)\right)$}
            \\ \hline
    \end{tabular}}}
    \caption{\small Ac\'in decomposition of the states of 3 qubits $\{\ket{M^{-1}_3}_h, \ket{M^{1}_3}_h, \ket{V^{-1}_3}_h, \ket{V^{1}_3}_h\}$ (where $j=1/2$). In this table the matrices $U_1,U_2,U_3$ such that $\ket{\tilde \Psi}_h = U_1 \otimes U_2 \otimes U_3\ket{\Psi}_h$ are collected as well as the coefficients $\lambda_0,\lambda_1,\lambda_2, \lambda_3, \varphi$ of {Eq}. \eqref{Acin}. {Note that we have chosen to work in the principal branch of the logarithm in the definition of $\varphi$}. $\ket{\tilde \Psi}_h$ corresponds to the Ac\'in decomposition of the state $\ket{\Psi}_h$. Notice that coefficients ${d_\pm}$ are given in Eq. \eqref{dcoef}. }
    \label{table1:acin}
\end{table}
\begin{table}[H]
    \centering
   \resizebox{0.8\textwidth}{0.8\textheight}{\Rotatebox{90}{ \begin{tabular}{|c||c|c|}
       \hline

            & $\ket{D^{-3}_3}_h$
            & $\ket{D^{-1}_3}_h$
        \\ 
        \hline\hline
        $U_1$ 
          
            & {$\left(
\begin{array}{cc}
 \frac{2}{\sqrt{\left(4 \sqrt{3}+7\right) h^2+4}} & \frac{\left(\sqrt{3}+2\right) h}{\sqrt{\left(4 \sqrt{3}+7\right) h^2+4}} \\
 -\frac{\left(\sqrt{3}+2\right) h}{\sqrt{\left(4 \sqrt{3}+7\right) h^2+4} \text{ sgn}\left(4 h-3 h^3\right)} & \frac{2}{\sqrt{\left(4 \sqrt{3}+7\right) h^2+4} \text{ sgn}\left(4 h-3 h^3\right)} \\
\end{array}
\right)$}
            &{ $\left(
\begin{array}{cc}
 -\frac{2}{\sqrt{\left(4 \sqrt{7}+11\right) h^2+4}} & -\frac{\left(\sqrt{7}+2\right) h}{\sqrt{\left(4 \sqrt{7}+11\right) h^2+4}} \\
 -\frac{\left(\sqrt{7}+2\right) h \text{\,sgn}\left(h \left(3 h^2-4\right) \left(7 h^2-4\right)\right)}{\sqrt{\left(4 \sqrt{7}+11\right) h^2+4}} & \frac{2 \text{ sgn}\left(h \left(3 h^2-4\right) \left(7 h^2-4\right)\right)}{\sqrt{\left(4 \sqrt{7}+11\right) h^2+4}} \\
\end{array}
\right)$}
        \\ \hline
        $U_2$ 
           
            & {$\left(
\begin{array}{cc}
 \frac{h}{\sqrt{h^2+\frac{4}{3}}} & \frac{2}{\sqrt{3 h^2+4}} \\
 -\frac{2 \text{ sgn}\left(4 h-3 h^3\right)}{\sqrt{3 h^2+4}} & \frac{h \text{ sgn}\left(4 h-3 h^3\right)}{\sqrt{h^2+\frac{4}{3}}} \\
\end{array}
\right)$}
            &{ $\left(
\begin{array}{cc}
 -\frac{h}{\sqrt{h^2+\frac{4}{7}}} & -\frac{2}{\sqrt{7 h^2+4}} \\
 -\frac{2 \text{ sgn}\left(h \left(7 h^2-4\right)\right)}{\sqrt{7 h^2+4}} & \frac{h \text{ sgn}\left(h \left(7 h^2-4\right)\right)}{\sqrt{h^2+\frac{4}{7}}} \\
\end{array}
\right)$ }
        \\ \hline
        $U_3$ 
        
            & {$\left(
\begin{array}{cc}
 \frac{h}{\sqrt{h^2-16 \sqrt{3}+28} \text{ sgn}(h)} & \frac{2}{\left(\sqrt{3}+2\right) \sqrt{h^2-16 \sqrt{3}+28} \text{ sgn}(h)} \\
 -\frac{2}{\left(\sqrt{3}+2\right) \sqrt{h^2-16 \sqrt{3}+28}} & \frac{h}{\sqrt{h^2-16 \sqrt{3}+28}} \\
\end{array}
\right)$}
            & {$\left(
\begin{array}{cc}
 \frac{h}{\sqrt{h^2+\frac{4}{4 \sqrt{7}+11}}} & \frac{2}{\left(\sqrt{7}+2\right) \sqrt{h^2+\frac{4}{4 \sqrt{7}+11}}} \\
 -\frac{2 \text{\,sgn}\left(h \left(4-3 h^2\right)\right)}{\left(\sqrt{7}+2\right) \sqrt{h^2+\frac{4}{4 \sqrt{7}+11}}} & \frac{h \text{\,sgn}\left(h \left(4-3 h^2\right)\right)}{\sqrt{h^2+\frac{4}{4 \sqrt{7}+11}}} \\
\end{array}
\right)$ }
        \\ \hline
        $\lambda_0$ 
           
            & {$ \sqrt{\frac{9 h^2+12}{19 h^4+32 h^2+16}} | h| $}
            & {$2 \sqrt{\frac{7 h^2+4}{9 h^4+32 h^2+48}}$}
        \\ \hline
         $\lambda_1$ 
             
            & {$\frac{2 \left(h^2+4\right) }{\sqrt{\left(3 h^2+4\right) \left(19 h^4+32 h^2+16\right)}}$} 
            & {$\frac{\sqrt{7}  | h|  \left| 4-3 h^2\right| }{\sqrt{\left(7 h^2+4\right) \left(9 h^4+32 h^2+48\right)}}$}
        \\ \hline
         $\lambda_2$ 
            
            & {$\frac{\sqrt{3} | h|  \left| 4-3 h^2\right| }{\sqrt{\left(3 h^2+4\right) \left(19 h^4+32 h^2+16\right)}}$}
            & {$\frac{2 \left| 4-7 h^2\right| }{\sqrt{\left(7 h^2+4\right) \left(9 h^4+32 h^2+48\right)}}$}
        \\ \hline
         $\lambda_3$ 
            
           & {$\frac{\sqrt{3} \left(h^2+4\right) | h| }{\sqrt{\left(3 h^2+4\right) \left(19 h^4+32 h^2+16\right)}}$}
            & {$\frac{2 \left| 4-3 h^2\right| }{\sqrt{\left(7 h^2+4\right) \left(9 h^4+32 h^2+48\right)}}$} 
        \\ \hline
         $\lambda_4$ 
            
           & {$\frac{12 h^2}{\sqrt{\left(3 h^2+4\right) \left(19 h^4+32 h^2+16\right)}}$}
            & {$\frac{8 \left| \left(11 \sqrt{7}+28\right) h^3+4 \sqrt{7} h\right| }{\left(\left(4 \sqrt{7}+11\right) h^2+4\right) \sqrt{\left(7 h^2+4\right) \left(9 h^4+32 h^2+48\right)}}$}
        \\ \hline
         $\varphi$ & {$-i \log_{\pi} \left(\text{ sgn}(4-3h^2) \right)$}& {$-i\log_{\pi} \left(\text{ sgn}\left(7 h^2-4\right)\right)$}
            \\ \hline
    \end{tabular}}}
    \caption{\small Ac\'in decomposition of the states of 3 qubits $\{\ket{D^{-3}_3}_h, \ket{D^{-1}_3}_h\}$ (characterized by $j=3/2$). In this table the matrices $U_1,U_2,U_3$ such that $\ket{\tilde \Psi}_h = U_1 \otimes U_2 \otimes U_3\ket{\Psi}_h$ are collected as well as the coefficients $\lambda_0,\lambda_1,\lambda_2, \lambda_3, \varphi$ of {Eq}. \eqref{Acin}. {Note that we have chosen to work in the principal branch of the logarithm in the definition of $\varphi$}.  $\ket{\tilde \Psi}_h$ corresponds to the Ac\'in decomposition of the state $\ket{\Psi}_h$. 
}
    \label{table2:acin}
\end{table}
\renewcommand{\arraystretch}{1}
\section{Clebsch--Gordan coefficients for 4-qubit \texorpdfstring{$h$}{h}-deformed states}\label{AppB}
\renewcommand{\arraystretch}{1.8}
\begin{table}[H]
\centering
\scriptsize
\resizebox{0.6\textwidth}{0.85\textheight}{\Rotatebox{90}{%
 \begin{tabular}{|| c || c | c | c | c | c | c | c | c | c | c | c | c | c | c | c | c ||} 
 \hline
 $N=4$ & $\ket{ \down\down \down\down}$ &$\ket{ \up\down \down\down}$ & $\ket{ \down\up\down \down}$ & $\ket{\down\down \up\down}$ & $\ket{ \down\down \down \up}$ &  $\ket{\up \up\down \down}$ & $\ket{ \up\down \up\down}$ & $\ket{ \up\down \down \up}$ &$\ket{\down \up \up \down }$ & $\ket{ \down\up\down \up}$ & $\ket{\down\down \up\up}$ & $\ket{ \up \up \up \down}$ &  $\ket{\up \up\down \up}$ & $\ket{ \up\down \up\up}$ & $\ket{ \down \up \up \up}$ &$\ket{ \up \up \up \up}$\\ 
 \hline\hline
$D^{-4}_4$& $1$ &$-3h/2$ & $-h/2$ & $h/2$& $3h/2$& $7h^2/4$ & $h^2/4$ & $-5h^2/4$ & $3h^2/4$& $h^2/4$ &$ 7h^2/4$ & $-9h^3/8$ & $5h^3/$8 & $-5h^3/8$& $9h^3/8$ & $9h^4/16$\\[1ex]
$D^{-2}_4$& 0 & $1/2$& $1/2$ & $1/2$ & $1/2$&$-h$ & $-h/2$& 0& 0&$h/2$ & $h$&$9h^2/8$ & $h^2/8$ & $h^2/8$&$9h^2/8$ &0 \\[1ex]
$D^{0}_4$& 0& 0& 0&0 & 0& $1/\sqrt{6}$& $1/\sqrt{6}$& $1/\sqrt{6}$& $1/\sqrt{6}$ &$1/\sqrt{6}$ & $1/\sqrt{6}$ &$ -h\sqrt{3/8}$ & $-h/\sqrt{24}$ &$h/\sqrt{24}$ & $ h\sqrt{3/8}$  & $h^2 \sqrt{3/8}$\\[1ex]
$D^{2}_4$&0 &0 &0 &0 & 0&0 &0 & 0& 0& 0&0 & 1/2& 1/2 & 1/2 & 1/2 &0 \\[1ex]
$D^{4}_4$&0 & 0& 0& 0& 0& 0&0 & 0& 0&0 & 0& 0& 0& 0& 0& 1\\[1ex]\hline
$M^{-2}_4$&0 &$\sqrt{3}/6$ & $\sqrt{3}/6$ & $\sqrt{3}/6$& $-\sqrt{3}/2$ & $-h/\sqrt{3}$& $-h \sqrt{3}/6$ &$h\sqrt{3}/3$ & 0&$-h\sqrt{3}/6$ & $-2h \sqrt{3}/3$ & $3h^2\sqrt{3}/8$&$-h^2\sqrt{3}/8$ & $5h^2\sqrt{3}/24$& $-11h^2 \sqrt{3}/24$& $-h^3\sqrt{3}/4$\\[1ex]
$M^{0}_4$& 0& 0& 0& 0& 0& $1/\sqrt{6}$&$1/\sqrt{6}$ &$-1/\sqrt{6}$ & $1/\sqrt{6}$& $-1/\sqrt{6}$ &$-1/\sqrt{6}$ &$-h\sqrt{6}/4$ &$-h/(2\sqrt{6})$ &$-h\sqrt{6}/4$ & $-5h/(2\sqrt{6})$& 0\\[1ex]
$M^{2}_4$& 0& 0& 0&0 & 0&0 & 0& 0& 0& 0& 0& $\sqrt{3}/2$& $-\sqrt{3}/6$ & $-\sqrt{3}/6$ & $-\sqrt{3}/6$ & $-h\sqrt{3}$\\[1ex]\hline
$V^{-2}_4$&0 & $1/\sqrt{6}$ &$ 1/\sqrt{6}$ & $-\sqrt{6}/3$& 0& $-h\sqrt{6}/3$ & $h\sqrt{6}/12$& $h\sqrt{6}/12$& $-h\sqrt{6}/4$ & $h\sqrt{6}/12$& $-h/\sqrt{6}$ & $h^2 \sqrt{6}/4$ &0 & $-h^2 \sqrt{6}/24$& $-5h^2 \sqrt{6}/24$ & $-h^3 \sqrt{6}/8$\\[1ex]
$V^{0}_4$& 0& 0& 0& 0& 0& $1/\sqrt{3}$& $ -\sqrt{3}/6$& $\sqrt{3}/6$ & $-\sqrt{3}/6$& $\sqrt{3}/6$ & $-1/\sqrt{3}$ & $-h\sqrt{3}/2$& $-h\sqrt{3}/6$ & 0& $-h\sqrt{3}/3$& 0\\[1ex]
$V^{2}_4$&0 & 0&0 & 0& 0&0 &0 & 0&0 &0 & 0& 0& $\sqrt{6}/3$& $-1/\sqrt{6}$ &$-1/\sqrt{6}$ & $-h\sqrt{6}/2$\\[1ex]\hline
$T^0_4$& 0& 0&0 & 0& 0& $1/\sqrt{3}$& $-\sqrt{3}/6$ & $-\sqrt{3}/6$ & $- \sqrt{3}/6$ & $-\sqrt{3}/6$&  $1/\sqrt{3}$& $-h \sqrt{3}/2$ & $-h \sqrt{3}/6$& $h\sqrt{3}/6$& $h\sqrt{3}/2$&$h^2\sqrt{3}/2$ \\[1ex]\hline
$R^{-2}_4$&0 & $1/\sqrt{2}$& $-1/\sqrt{2}$&0 & 0& $-h/\sqrt{2}$& $-h/\sqrt{8}$ &  $h/\sqrt{8}$& $h/\sqrt{8}$ & $-h\sqrt{8}$ & 0& $h^2/\sqrt{8}$& $-h^2/\sqrt{8}$ & $h^2/\sqrt{32}$&  $-h^2/\sqrt{32}$& $-h^3/\sqrt{32}$ \\[1ex]
$R^0_4$&0 & 0& 0&0 & 0& 0& $1/2$ &  $1/2$& $-1/2$ & $-1/2$ & 0& $-h/2$& $-h/2$ & 0& 0 &0 \\[1ex]
$R^2_4$&0 & 0& 0&0 & 0& 0& 0 &  0& 0 & 0 & 0& 0& 0 & $1/\sqrt{2}$&  $-1/\sqrt{2}$& $-h/\sqrt{2}$ \\[1ex]\hline
$Q^0_4$&0 & 0& 0&0 & 0& 0& $1/2$ &  $-1/2$& $-1/2$ & $1/2$& 0& $-h/2$& $h/2$ & $-h/2$ & $h/2$  & $h^2/2$\\[1ex] 
 \hline
 \end{tabular}}}
 \normalsize
 \caption{\small Explicit expressions for {non-normalized} $h$-states with $4$ qubits as linear combinations of the elements of the basic/computational basis (in the first row of the table). The coefficients were obtained by iteratively using the corresponding Clebsch--Gordan transformations. }
  \label{table:4q}
 \end{table}
 \renewcommand{\arraystretch}{1}

\section{Proof of \texorpdfstring{Eq.~\eqref{eq:main_identity}}{eq}}
\label{appendix_proof}
The objective of this Appendix is to prove Eq.~\eqref{eq:main_identity}. Let  $P=\{x_1,x_2,\dots,x_n\}$ a set with cardinal $n$. For any integers $r, i$ such that $r \le n-i$, the following identity holds:
\begin{equation}\label{eq67}
    \sum_{J \subset P, |J|=r} e_i(P \setminus J) = \binom{n-i}{r} e_i(P),
\end{equation}
where $J$ is any subset of $P$ with $r$ elements. 

This proof starts from the following fundamental recurrence relation for elementary symmetric polynomials in the case $r=1$ \cite{ELMIKKAWY20138770, Macdonaldbook}:
\begin{equation}
    e_i(P \setminus \{x_k\}) = e_i(P) - x_k \cdot e_{i-1}(P \setminus \{x_k\}).
    \label{eq:recurrence}
\end{equation}
This relation is valid because the terms of $e_i(P)$ can be separated into those that do not contain $x_k$ (which form $e_i(P \setminus \{x_k\})$) and those that do (which can be written as $x_k$ multiplied by the terms of $e_{i-1}(P \setminus \{x_k\})$).

We will proceed by induction on $r$. First, we prove the identity for $r=1$. We want to show that 
\begin{equation}
	\sum_{k=1}^n e_i(P \setminus \{x_k\}) = \binom{n-i}{1} e_i(P) = (n-i)e_i(P).
\end{equation}
Summing Eq.~\eqref{eq:recurrence} over all $x_k \in \{1, \dots, n\}$:
$$ \sum_{k=1}^n e_i(P \setminus \{x_k\}) = \sum_{k=1}^n e_i(P) - \sum_{k=1}^n x_k \cdot e_{i-1}(P \setminus \{x_k\}) $$
The first term on the right is simply $n \cdot e_i(P)$. The second term, $\sum_{k=1}^n x_k  \cdot e_{i-1}(P \setminus \{x_k\})$, is the sum of all products of a variable $x_k$ with a monomial of degree $i-1$ that does not contain $x_k$. This means that multiplying such a monomial by $x_k$ gives all degree $i$ monomials that contain $x_k$, and only those monomials. Each monomial of degree $i$ (e.g., $x_{j_1} \cdots x_{j_i}$) appears exactly $i$ times in the total sum, exactly once for each of its variables. Therefore
\begin{equation}
	\sum_{k=1}^n e_i(P \setminus \{x_k\}) = n \cdot e_i(P) - i \cdot e_i(P) = (n-i)e_i(P) .
\label{eq:medium}
\end{equation}
The base case has been proven.

Let us start the proof by induction. We assume that identity \eqref{eq67} holds for subsets of size $r-1$:
\begin{equation}
	\sum_{J' \subset P, |J'|=r-1} e_i(P \setminus J') = \binom{n-i}{r-1} e_i(P) .
\end{equation}
Now, we apply the double sum method as an intermediate tool in the inductive proof because it allows us to count the same set of terms in two different ways, giving the relationship between the sum we want to compute and the sum over smaller subsets, to which one can apply the inductive hypothesis. Hence, we consider the following summation over subsets of $r$ elements:
\begin{equation}\label{eq72}
	\sum_{J \subset P, |J|=r} \sum_{x_j \in J} e_i(P \setminus J).
\end{equation}
We will evaluate it in two different ways, following different orders. First, we fix the subset $J$ of size $r$, and for each of its $r$ elements, we add the same value $e_i(P \setminus J)$. The inner sum over $x_j \in J$ has $r$ identical terms. Therefore:
\begin{equation}\label{eq73}
	\sum_{J \subset P, |J|=r} \sum_{x_j \in J} e_i(P \setminus J) = r \sum_{J \subset P, |J|=r} e_i(P \setminus J).
\end{equation}

In the second method of evaluating the sum, we swap the order of the summation. We first fix the element $x_j$ and then sum over all subsets $J$ of size $r$ that contain $x_j$. Any subset $J$ of size $r$ that contains $x_j$ can be written as ${x_j} \cup J'$, where $J'$ is a subset of $P \setminus {x_j}$ of size $r-1$. Thus, we want to sum
\begin{equation}\label{eq74}
	\sum_{x_j \in P} \sum_{\substack{J' \subset P \setminus \{x_j\} \\ |J'|=r-1}} e_i(P \setminus (J' \cup \{x_j\})).
\end{equation}
The term $P \setminus (J' \cup \{x_j\})$ is equal to $(P \setminus \{x_j\}) \setminus J'$. Therefore, the inner sum is a sum over subsets of size $r-1$ of a set of size $n-1$ (the set $P \setminus \{x_j\}$). Applying our inductive hypothesis to this set:
\begin{equation}
	\sum_{\substack{J' \subset P \setminus \{x_j\} \\ |J'|=r-1}} e_i((P \setminus \{x_j\}) \setminus J') = \binom{(n-1)-i}{r-1} e_i(P \setminus \{x_j\}).
\end{equation}
Substituting this into the double sum \eqref{eq74} yields:
\begin{equation}
	\sum_{x_j \in P} \binom{n-i-1}{r-1} e_i(P \setminus \{x_j\}) = \binom{n-i-1}{r-1} \sum_{x_j \in P} e_i(P \setminus \{x_j\}) .
\end{equation}
However, the summation on the right-hand side runs over subsets of $P$ with only one element, so by using Eq.~\eqref{eq:medium}, we know that:
\begin{equation}\label{eq78}
	\binom{n-i-1}{r-1} \sum_{x_j \in P} e_i(P \setminus \{x_j\}) = \binom{n-i-1}{r-1} (n-i) e_i(P).
\end{equation}
Now we equate the two ways (eq. \eqref{eq78} and \eqref{eq73}) we evaluated the double sum \eqref{eq72}:
\begin{equation}
	 r \sum_{J \subset P, |J|=r} e_i(P \setminus J) = \binom{n-i-1}{r-1} (n-i) e_i(P).
\end{equation}
By isolating the sum of interest we find
\begin{equation}
	\sum_{J \subset P, |J|=r} e_i(P \setminus J) = \frac{n-i}{r} \binom{n-i-1}{r-1} e_i(P),
\end{equation}
which, after using the well-known properties for binomial coefficients reduces to the desired result
\begin{equation}
	\sum_{J \subset P, |J|=r} e_i(P \setminus J) = \binom{n-i}{r} e_i(P).
\end{equation}

\end{document}